\documentclass[5p]{elsarticle} %

\input{usepackages.sty}
\input{newcommands.sty}

\makenomenclature

\makeindex

\begin{document}

\title{Characteristic Methods and Roe's Method for the Incompressible Two-Fluid Model for Stratified Pipe Flow  }

\author{A.H.~Akselsen\corref{cor1}}
\address{Department of Energy and Process Engineering, Norwegian University of Science and Technology, Kolbj{\o}rn Hejes v.\ 1B, 7491 Trondheim, Norway}
\ead{andreas.h.akselsen@ntnu.no}

\cortext[cor1]{Corresponding author}

\begin{keyword}
two-phase; pipe-flow; two-fluid model; method of characteristics; Roe scheme
\end{keyword}%

\renewcommand{\today}{August 30, 2016}
 
\begin{abstract}
This article examines the use of characteristic methods in stratified two-phase pipe flow simulations for obtaining non-dissipative flow predictions. 
A Roe scheme and several methods based on the principle of characteristics are presented for the two-fluid model. 
The main focus is finding numerically efficient ways of capturing wave dynamics and flow regime transitions through direct simulation. 
Characteristic methods offer the possibility of simulating hyperbolic systems without numerical dissipation.
These methods do however lack certain fundamental conservation properties.
Challenges related to information scattering and clustering in space and time, particularly around shocks, are also an issue  in some method variants.
Hybridisations with the finite volume method are proposed which overcome these shortcomings.
All methods are compared, evaluating predictions on the onset of linear wave growth and simulations of non-linear, discontinuous  roll-waves.
The following observations are made:
1. Characteristic methods are excellent at predicting the onset of linear hydrodynamic instability, even with a small number of computational nodes. 
2. Dissipative errors in finite volume methods and characteristic hybrids will be closely linked to the \Courant{} number. The Roe scheme and the characteristic hybrids give very little dissipation error as the \Courant{} number approaches unity. This then becomes a question of numerical stability. 
3. Adapting dynamic grid cells, moving along with the characteristics information drift, greatly improves simulation efficiency 
by allowing for longer time steps. Dynamic grid cells are also useful for alleviating the need for interpolation in characteristic methods. 

On the whole, the performance of the characteristic hybridisations are similar to that of the Roe scheme at large \Courant{} numbers. Characteristic hybridisations perform somewhat better in predicting linear instability while the Roe scheme is better suited for the shock fronts found in the roll-wave flow regime. 
These method easily out-perform the more basic upwind and Lax-Friedrich schemes.

\end{abstract}
\maketitle

\section{Introduction}
The method of characteristics, henceforth abbreviated `MOC', has been common amongst hydraulic engineers many years, particularly for calculations regarding sonic waves in long pipelines (waterhammer) \cite{Streeter_MOC_numerical_methods, Streeter_waterhammer_unread}.
These are problems {of weakly compressible single-phase flows} where the sonic information travels very fast in both directions compared to the convective velocity.
The MOC does not suffer from numerical diffusion in the same way that finite difference, volume or element methods do, which makes it suitable for expressing the long-range sonic waves of pipe systems. 

The method has also been used for \textit{multiphase} pipe flow problems, such as for dispersed flows \cite{Bournaski_MOC_dispersed_virtual_mass}.
The non-dissipative nature of the method also makes it attractive in relation to surface wave phenomena. 
{Crowley et.\ al.~\cite{Crowley_MOC_VKH_1991} and} Barnea and Taitel \cite{Barnea_simplified_with_characteristics,Barnea_stability_separated_flow}
simulated stratified two-phase pipe flow {under simplifying assumptions} using the MOC. 
They compared these simulation results to analytical expressions from linear stability theory.
The method was here chosen so that numerical diffusion would not artificially stabilise the flow. 

{Volume waves} are a vital element in the evolution of stratified pipe flows. 
Characteristics methods are however uncommon in multiphase pipe flow simulation software, the reasons for which can by and large be boiled down to three shortcomings in the MOC: 
\begin{enumerate}[1.]
\itemsep0pt
	\item The MOC does not provide a numerically conservative formulation. Numerical errors may then accumulate and manifest in the false appearance or disappearance of mass, momentum, energy, etc..
	\item The distribution  of numerical  nodes in space and time turns irregular if the problem is strongly non-linear. 
	\item The method is generally not well-suited for handling discontinuities. 
\end{enumerate}

More common for simulating non-linear hyperbolic problems are Roe's approximate Riemann solvers~\cite{Roe_original}, which are designed to provide solutions to a linearised shock problem.

Toumi and Kumbaro \cite{Toumi_Roe_homogenious_two_fluid_model,Toumi_Riemann_solver_two_fluid_model} were amongst the first to formulate Roe schemes for compressible two-phase pipe flow models. This flow models represented by these schemes cannot be put in a conservative form; providing weak formulations of the non-conservative terms is a central feature of this work.

Fl{\aa}tten and Munkejord~\cite{Munkejord_Roe_on_drift_flux} presented a strategy for constructing a Roe-average matrix for the so-called drift-flux model for two-phase pipe flows. The strategy did not place any requirements on the drift flux function -- it need not even be an algebraic expression. 

A last method worth mentioning for simulating stratified two-phase pipe flow is Holm{\aa}s' pseudospectral scheme~\cite{Holmaas_roll_wave_model}. Spectral methods are not included here as their application to non-periodic boundary problems in unclear.
\\

The objective of this paper is the study and development of simple and efficient methods for simulating stratified pipe flows, focusing primarily on characteristic methods. 
An efficient method should be capable of both detecting the onset of hydrodynamic wave growth and the wavy flow regime that ensues. 
Therefore, methods are tested up against analytical expressions for the onset of linear instability~\cite{Barnea_stability_separated_flow} and discontinuous roll-waves~\cite{Watson_wavy}.

The structure of this article is as follows:
The incompressible two-fluid model for stratified pipe flow is presented in \autoref{sec:two_fluid_model}.
Theory for the construction of characteristic methods and finite volume methods are presented in \autoref{sec:MOC_theory} and \ref{sec:averaged_eqs}, respectively. A number of characteristics-based methods are presented in \autoref{sec:all_MOC_based}, sequentially focusing on remedying some of the issues associated with the characteristics approach. 
A Roe method has been derived and is presented in \autoref{sec:Roe}.
These methods are then compared through the numerical experiments of \autoref{sec:exp}, where the methods are tested in both the linear and non-linear wave regimes. 
Discussion and conclusions are given in \autoref{sec:discussion} and \ref{sec:conclusions}, respectively.

\section{The Two-Fluid Model for Pipe Flow}
\label{sec:two_fluid_model}

\begin{figure}[h!ptb]
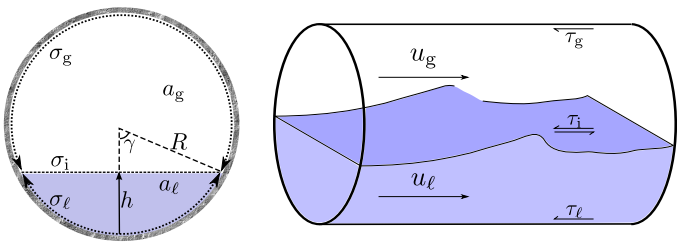
%
\centering
\begin{minipage}[c]{.35\columnwidth}
\includegraphics[width=\columnwidth]{./images/pipe_cross_section3.\prykkexp}%
\end{minipage}
\hfill
\begin{minipage}[c]{.6\columnwidth}
\includegraphics[width=\columnwidth]{./images/pipe_sideways2.\prykkexp}%
\end{minipage}
\caption{Pipe cross-section}%
\label{fig:cross_section}%
\end{figure}
\autoref{fig:cross_section} illustrates the pipe geometry and some of the quantities appearing the two-fluid model.
Field $\phaseindex$, occupied by either gas $\phaseindex = \mr g$ or liquid $\phaseindex = \ell$, is segregated from the other field. Subscript $\mr i$ indicates the fluid interface.
The circular pipe geometry itself enters into the modelling through the relation between the level height $h$, the specific areas $\ak$ and the perimeter lengths $\sigma\k$ and $\sigma\_i$.
These are algebraically interchangeable through a geometric function
\begin{equation}
h = \Hofal\of \al.
\label{eq:Alofh}
\end{equation}
Only the inverse of $\Hofal$ is an explicit expression
\begin{align*}
\Hofal\inv\of h &= R^2\br{\gamma - \tfrac12 \sin 2\gamma},
&
\gamma\of h &= \arccos\br{1-\frac h R},
\end{align*}
but Biberg's approximation \cite{Biberg_h_of_al}
\begin{align*}
\Hofal\of \al  &= R \br{1-\cos\gamma},
\\
\gamma & \approx\ \pi \alpha\l + \br{\tfrac{3}{2}\pi}^{\!\!\nicefrac13}\!\br{1-2\alpha\l+\alpha\l^{\nicefrac13}-\alpha\g^{\nicefrac13}}\\
&\quad\;\; -0.005\,\alpha\l\alpha\g\br{\alpha\g-\alpha\l}\br{1+4\br{\alpha\l^2+\alpha\g^2}^2},
\end{align*}
is very accurate.
$R$ is here the pipe inner radius, $\gamma$ the interface half-angle and $\alpha\k = \ak/\area$ is the phase fraction.
The function derivative is $\dHofal = 1/\sigma\_i$
and the perimeter lengths are 
\begin{align}
\sigma\l &= 2 R \gamma,
&
\sigma\g &= 2R\br{\pi-\gamma},
&
\sigma\_i &= 2R\sin \gamma.
\label{eq:sigma}
\end{align}

The compressible, adiabatic, equal pressure four-equation two-fluid model for stratified pipe flow results from an averaging of the conservation equations across the cross-section area.  
The model is commonly written
\begin{subequations}
\begin{align}
	&\br{\rho\k  a\k }_t+  \brac{\rho\k  a\k  u\k}_x = s_k\^{mass}, \label{eq:base_model_compressible:mass} \\
	& \br{\rho\k   a\k  u\k}_t + \brac{\rho\k  a\k  u\k^2}_x +  a\k p_{\mr i, x} + \rhok \ak g_y    h_x  = s_k\^{mom}. 
\end{align}
\label{eq:base_model_compressible}
\end{subequations}
$ p\_i$ is here the pressure at the interface, assumed the same for each phase as surface tension is neglected.
$h$ is the height of the interface from the pipe floor, and the term in which it appears originates from approximating a hydrostatic wall-normal pressure distribution within a fluid. 
$u\k$ and $\rho\k$ are the mean fluid velocity and density in field $\phaseindex$. 
The momentum sources are 
$
s_k\^{mom} = \tau\k \sigma\k \pm \tau\_i \sigma\_i - \rho\k\a\k  g_x, 
$
where $\tau$ is the skin frictions at the walls and interface.
{$g_x$ and $g_y$ are respectively the axial and transverse components of the gravitational acceleration, \ie, $g_x=g \sin \theta$ and  $g_y=g \cos \theta$ if} $\theta$ is the pipe inclination, positive above datum.
Internal mass sources $s_k\^{mass}$ are commonly zero.

Both fluid flows are from here assumed incompressible. 
{Assuming incompressible phases entails that acoustic waves travel at infinite speeds. 
Acoustic waves are thus eliminated from the problem, manifesting instead in an algebraic manner.
The pressure then ceases to function as a transport variable and it may be eliminated from the system, allowing \eqref{eq:base_model_compressible} to be written in a mathematically conservative two-equation form. This conservative form is significantly easier to handle from a numerical perspective.}

The pressure is eliminated by reducing the momentum equations with their respective mass equations, dividing by the respective phase areas and  differencing the two, yielding
\begin{equation}
	\bw_t + \bff_x = \bm s
\label{eq:base_model}
\end{equation}
where
\begin{subequations}
\begin{align}
\bw &= 
\begin{pmatrix}
	 a\l \\ \q 
\end{pmatrix},
\\
\bff &= 
\begin{pmatrix}
	\al u\l \\ \tfrac12 \rho\l  u\l^2- \tfrac12\rho\g \ug^2 +  \my h
\end{pmatrix}, \label{eq:f}
\\
\bm s &= 
\begin{pmatrix}
	s_{\al} \\ s_\q
\end{pmatrix},
\\
\q &=  \rho\l \ul-\rho\g\ug,
\\
s_\q &= -\mx -\frac{\tau\l \sigma\l}\al+\frac{\tau\g \sigma\g}\ag + \tau\_i \sigma\_i \br{\frac1\al+\frac1\ag}.
\label{eq:s}
\end{align}
\end{subequations}
Although actually derivatives therefrom, equations \eqref{eq:base_model} are here simply referred to as base mass and momentum equations.

The identities
\begin{align}
\al + \ag &= \area, & (\a u)\l+(\a u)\g &= \Qm,
\label{eq:sum_a_au}
\end{align}
where both right hand sides are parametric, finally close the base model \eqref{eq:base_model}.
{The cross-section pipe area $\area$ may vary spatially, but is kept constant in the presented examples.}
The second identity in \eqref{eq:sum_a_au} originates from summing the gas and liquid mass equations \eqref{eq:base_model_compressible:mass} and applying the first identity.
The 
{total} flow rate $\Qm$ is generally expressible as $\Qm = \mc Q\_{0} + \int_{x_0}^{x}\!(s_{a\g}+s_{a\l})\mr d\xi$.
Usually, internal mass sources $s_{a\k}\equiv0$ such that $\Qm$ everywhere equals the inlet mixture flow rate, made constant in all examples presented herein.

Primitive variables are recovered through
\begin{align*}
u\l &= \frac{ \rho\g \Qm +  \ag q}{\ag\rho\l+\al\rho\g},
&
u\g &= \frac{ \rho\l \Qm -  \al q}{\ag\rho\l+\al\rho\g}.
\end{align*}

\begin{figure}[h!ptb]%
\centering
\includegraphics[width=.5\columnwidth]{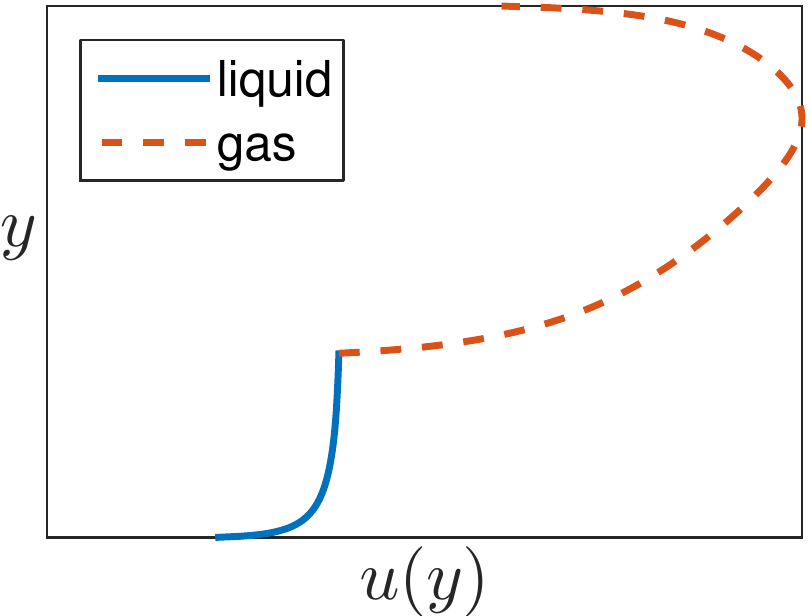}%
\caption{Example velocity profile pre-integrated in the Biberg friction model \cite{Biberg_friction_duct_2007}.  }%
\label{fig:Biberg_profile}%
\end{figure}

The friction closures $\tau\k$ and $\tau\_i$ in the numerical tests of this article are provided by the Biberg friction model as presented in \cite{Biberg_friction_duct_2007}. 
Here, classical turbulent  boundary layer principles are used to model the gas and liquid velocity profiles in a duct cross section. 
The interface is modelled as a moving boundary with an initial turbulence level.
\autoref{fig:Biberg_profile} shows an example of such a velocity profile. 
These profiles are integrated to yield algebraic expressions that couple wall and interfacial frictions to the average phase velocities and the interface height. 
Friction correlations for duct flow are correlated to the well-known Colebrook-White formula for single-phase pipe flow, which in turn is used to extend the friction model for two-phase duct flow into formulae for the pipe geometry.

\section{A Method of Characteristics}
\label{sec:MOC_theory}

The two-fluid model \eqref{eq:base_model} is expressed through a Jacobian $\Jac = \pdiff{\bff}{\bw}$ as follows:
\begin{equation}
\bw_t + \Jac \bw_x = \bs
\label{eq:base_model:J}
\end{equation}
where
\begin{equation}
\Jac = \frac{1}{\frac{\rho\l}{\al}+\frac{\rho\g}{\ag}}
\begin{pmatrix}
\frac{\rhol \ul}{\al}+\frac{\rhog \ug}{\ag} & 1\\
\kappa^2 & \frac{\rhol \ul}{\al}+\frac{\rhog \ug}{\ag}
\end{pmatrix}
\label{eq:Jac}
\end{equation}
and
\begin{equation}
	\kappa = \sqrt{\br{\frac{\rhol}{\al}+\frac{\rhog}{\ag}}\my\dHofal - \frac{\rho\l \rho\g}{\al \ag}\br{u\g-u\l}^2}.
\label{eq:kappa}
\end{equation} 
The eigenvalues $\lambda\ud\pm$ of $\Jac$, obtained from the roots of $\det(\Jac-\lambda \,\I)=0$, are
\begin{equation}
\lambda\ud\pm = \left.\br{\frac{\rhol \ul}{\al}+\frac{\rhog \ug}{\ag} \pm \kappa}\right/\br{\frac{\rho\l}{\al}+\frac{\rho\g}{\ag}}.
\label{eq:lambda}
\end{equation}

\begin{figure}[hptb]%
\centering
\includegraphics[width=.5\columnwidth]{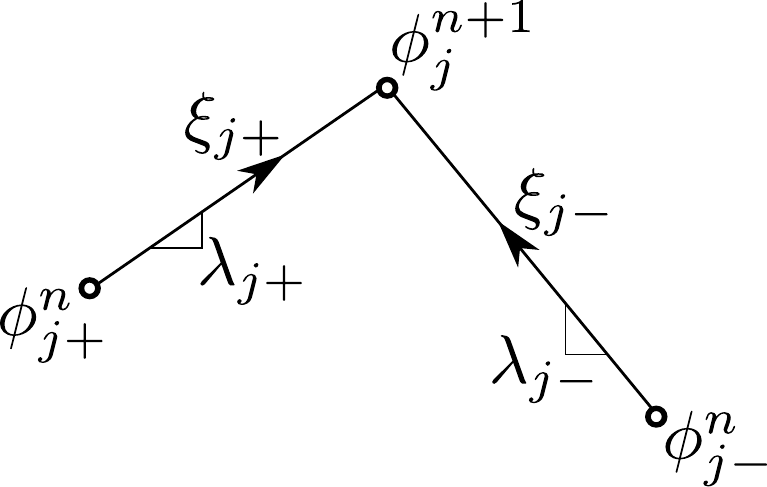}%
\caption{MOC index convention}%
\label{fig:MOC_path_notation}%
\end{figure}

System~\eqref{eq:base_model} will be hyperbolic if $\kappa$ is real.
Left eigenvectors $\Ll\ud\pm$ (row vectors) are used for the decomposition; 
\begin{equation}
\Jac = \LL\inv \Lamb \LL,
\label{eq:eigen_equation}
\end{equation}
where 
\begin{align}
	\LL &= 
	\begin{pmatrix}
		\Ll\ud+\\ \Ll\ud-
	\end{pmatrix},
	&
	\Ll\ud\pm &= 
	\begin{pmatrix}
		1&\pm 1/\kappa
	\end{pmatrix},
	&
	\Lamb &= 
		\begin{pmatrix}
		\lambda\ud+&0\\ 0&\lambda\ud-
	\end{pmatrix}.
	\label{eq:eigenvectors}
\end{align}
Inserting \eqref{eq:eigen_equation} into \eqref{eq:base_model:J} and multiplying by $\LL$ from the left gives an orthogonal system
\begin{equation}
\Ll\ud\pm \ddiff{\bw}{t} = \Ll\ud\pm\bs, 
\qquad \mr{along}~ \ddiff x t = \lambda\ud\pm.
\label{eq:diag_syst}
\end{equation}
{Note that, except for the gravitational component of $\kappa$, the Jacobian, eigenvalues and eigenvectors are symmetric with respect to the phases.}
The integral form of \eqref{eq:diag_syst} is
\begin{equation}
\alj\nn - \aljpm\n \pm \!\!\int\limits_{\q\jpm\n}^{\q_j\nn} \!\!\!\frac{\mr d \q}{\kappa}
= \int\limits_{t\jpm\n}^{t_j\nn} \!\!\!\br{s_{\al} \pm \frac{s_\q}{\kappa}} \mr d t,
\label{eq:diag_syst_int}
\end{equation}
where integration has been performed from $(x,t)\jpm\n$ to $(x,t)_\j\nn$ along the path $\ddiff x t = \lambda\ud\pm$. Subscripts $\jpmm$ indicate that state or point reached by following the $\pm$-path backwards in time from $(x,t)_\j\nn$.
{An illustration is shown in \autoref{fig:MOC_path_notation}.}
In discrete representations, the integration paths are linearised into line paths
\begin{equation}
\x\jpm\n\of t = x\jpm\n + \lambda\jpm\pmnh \br{t-t\jpm\n}, \quad t\jpm\n \leq t \leq t_j\nn.
\label{eq:x_of_t}
\end{equation}
with some intermediate slope $\lambda\jpm\pmnh$.

Also linearising equation \eqref{eq:diag_syst_int} along these respective paths gives an algebraic system 
\begin{equation}
\alj\nn - \aljpm\n \pm \frac{\q_j\nn-\q\jpm\n}{\kappa\jpm\nh} = 
\sqbr{\br{s_{\al} \pm \frac{s_{\q} }{\kappa}}\dt}\jpm\nh
\label{eq:diag_syst_disc}
\end{equation}
with 
$
\dt\jpm\n={t_j\nn-t\jpm\n}.
$ 
The state $\bw$ is obtainable only at the start and end of the integration, where characteristic paths intersect. 
Intermediate integrands $\phi\jpm\nh$ are therefore computed as some average of the available intersection states, depending on the method.

Solving \eqref{eq:diag_syst_disc} is straight forward. 
Suppressing the superscript $n+\frac12$, 
one finds
\begin{equation}
\bw_j\nn = \frac{
\begin{pmatrix}
	\kappa\jmm\inv & \kappa\jpp\inv \\
	1				&  -1
\end{pmatrix}
\begin{pmatrix}
	\br{\al\n + s_{\al}\dt\n + \frac{\q\n + s_{\q}\dt\n }{\kappa}}\jpp  \\
	\br{\al\n + s_{\al}\dt\n - \frac{\q\n + s_{\q}\dt\n }{\kappa}}\jmm
\end{pmatrix}
}{\kappa\jpp\inv + \kappa\jmm\inv}
.
\label{eq:MOC_end_eq}
\end{equation}

\section{Average Volume Equations for Dynamic Finite Volume Methods}
\label{sec:averaged_eqs}
{Finite volume methods have the advantage over characteristic methods of being conservative. 
For the purpose of hybridising these methods the  
control volumes presented here are allowed to move and stretch in time.
}

Integration of \eqref{eq:base_model} is performed across a grid cell $\J$, first in space from the left cell face at $x\L$ to the right one at $x\R$, and then in time from the present time $\tn$ to the next time level $\tnn$;
\begin{equation}
\int_{\tn}^{\tnn}\!\!\!\!\int_{x\L}^{x\R} \! \brac{ \partial_t \bw + \partial_x \bff - \bs}\,\mr dx\, \mr dt = 0.
\label{eq:Holmaas_integral}
\end{equation}
Using Leibniz' rule ($x\jpmh$ are here generally functions of $t$,) the first transient term evaluates to
\begin{equation*}
	\int_{x\L}^{x\R} \! \partial_t \bw \,\mr dx 
	= \pp_t \!\br{\dx \ol{\bw}} - \br{\cc \, \bw}\R+\br{\cc \, \bw}\L
\end{equation*}
where $\cc = \ddiff xt$ is the control volume border velocity.
$\dx \of t =  x\R -x\L $ is the cell length. 
The bar indicates the wave cell average and is defined
\begin{subequations}
\begin{equation}
	\ol\phi_\J\of{t} = \frac1{\dx } \int_{x\L}^{x\R} \! \phi\,\mr dx.
	\label{eq:averaging:space}
\end{equation}
Also introducing the temporal average $\an\cdot\n$, 
\begin{equation}
	\an{\phi}\n = \frac1{\dt\n}  \int_{\tn}^{\tnn} \!\!\!\!\!\!\!\! \phi\,\mr dt
	\label{eq:averaging:time}
\end{equation}
\end{subequations}
with $\dt\n = \tnn-\tn$, the integral equation \eqref{eq:Holmaas_integral}  is cast as
\begin{equation}
	\br{\dx \ol\bw}_\J\nn = \br{\dx \ol{\bw}}_\J\n +  \dt\n \avgt{ \dx \ol\bs_\J - \rel{\bff}\R+\rel{\bff}\L}, 
	\label{eq:avg_Holmaas}
\end{equation}
where the flux terms have been made relative to 
{the control volume border velocities $\cc\L$ and $\cc\R$. Dropping indexation, we have}
\begin{align}
\rel\bff &= \bff - \cc\bw 
\label{eq:fr}
\\&= 
\nonumber	\begin{pmatrix}
	\al (u\l-\cc) 
	\\ 
\rho\l \ul \br{\frac \ul2-\cc}-\rhog \ug \br{\frac \ug2-\cc} +  \my h
\end{pmatrix}.
\end{align}
It is important to note that \eqref{eq:avg_Holmaas} is still exact.
{
Straight border paths will be used in the schemes presented below such that all $\cc_j\n$ are constant during the time integration step. 
Border velocities can differ at each border and will be updated in between every time step, allowing control volumes to stretch and contract.
}

\section{Schemes Based on the Method of Characteristics}
\label{sec:all_MOC_based}

Seeking to overcome some of the weaknesses of data scattering and lacking conservation, a number of methods will here be suggested.
The first two methods are commonly found in the literature, both usually referred to as `the method of characteristics.' To differentiate between them, all methods are dubbed some alternation of `MOC.'
\\

A wide hat notation $\mhat \phi$ is from here added to
variables computed from \eqref{eq:MOC_end_eq} in order to better distinguish the MOC states of \autoref{sec:MOC_theory} from the averaged finite volume state variables $\ol \phi$ of  \autoref{sec:averaged_eqs}.

\subsection{A Method of Scattered Point Characteristics (\MOSC{} )}
\label{sec:MOC}
This method is similar to that used by {Crowley et.\ al.~\cite{Crowley_MOC_VKH_1991} and} Barnea and Taitel~\cite{Barnea_stability_separated_flow} for comparing the results from linear stability theory to simulations without numerical diffusion. 
The authors did however use a simplified version of model \eqref{eq:base_model} where the transient terms of the gas phase had been ignored.

\begin{figure}[h!tbp]%
\centering
\includegraphics[width=.8\columnwidth]{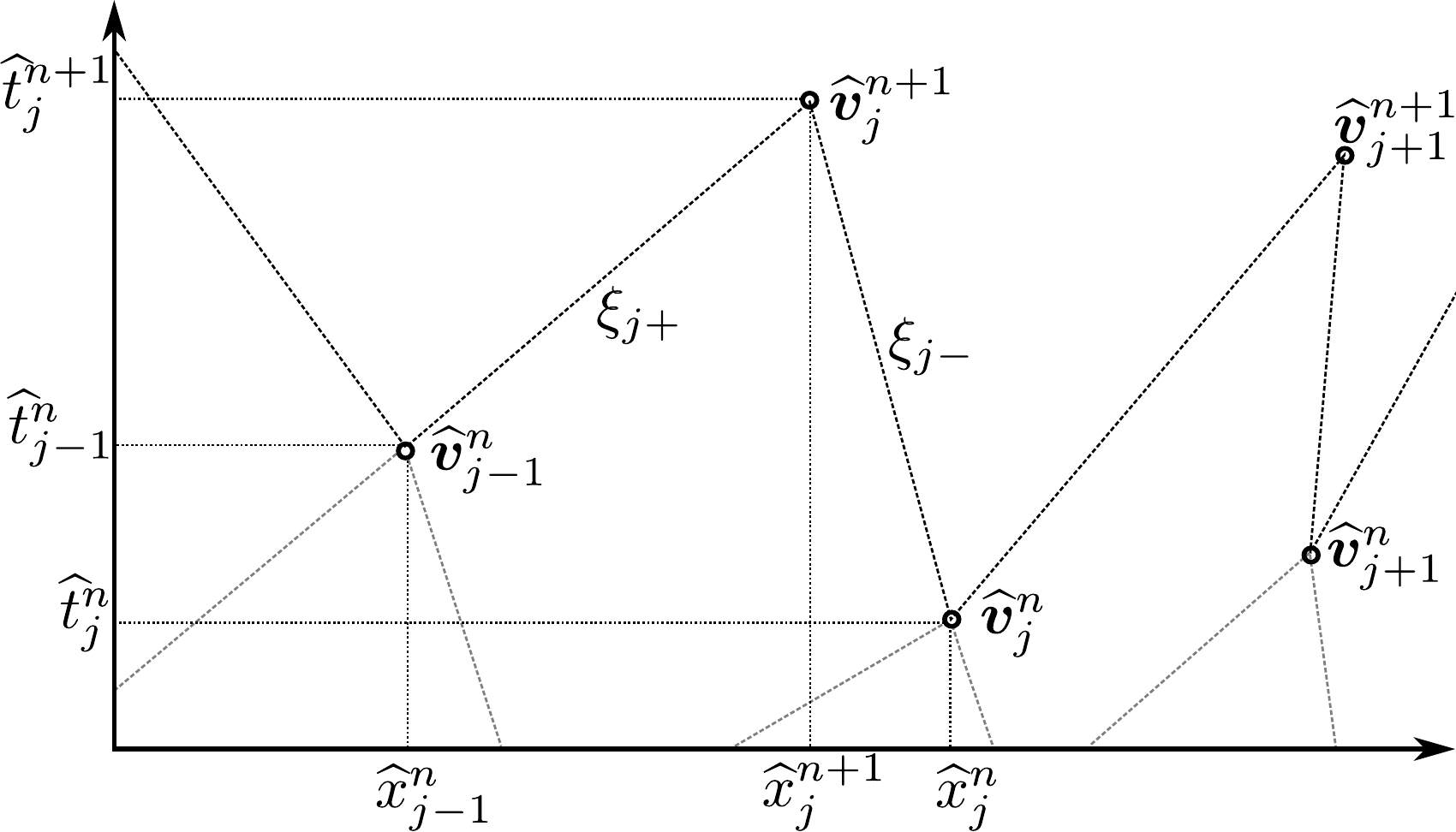}%
\caption{Sketch of the \MOSC{}.}%
\label{fig:MOC:scheme}%
\end{figure}

\autoref{fig:MOC:scheme} shows a schematic of the discrete Method Of Scattered Point Characteristics (\MOSC{}.)

The method consists of solving \eqref{eq:MOC_end_eq} at single-point locations in space and time where the characteristic paths intersect. 
With the presented indexing this means that $\mhat \bw\jpm\n = \mhat \bw_{\j\mp\frac12-\frac12}\n$.
Intersection points $(\mhat x,\mhat t)_\j\nn$ are found through \eqref{eq:x_of_t} by solving 
$
\mhat \x\jpm\n\of{\mhat t_\j\nn} = \mhat x_\j\nn.
$
Intermediate integrands are first chosen $\mhat \phi\jpm\nh=\phi\of{\mhat \bw\jpm\n}$ and are then iterated upon with
$\mhat \phi\jpm\nh =  \phi\of{\frac12\br{\mhat \bw\jpm\n+\mhat \bw_\j\nn}}$.
Iteration is not required, 
{but is sometimes worth doing to improve accuracy and stability -- it will be adopted when computing \eqref{eq:MOC_end_eq} in all numerical tests. 
A single iteration, at most two, can be worthwhile in the tests presented.}

\begin{figure}[h!ptb]%
\centering
\includegraphics[width=.5\columnwidth]{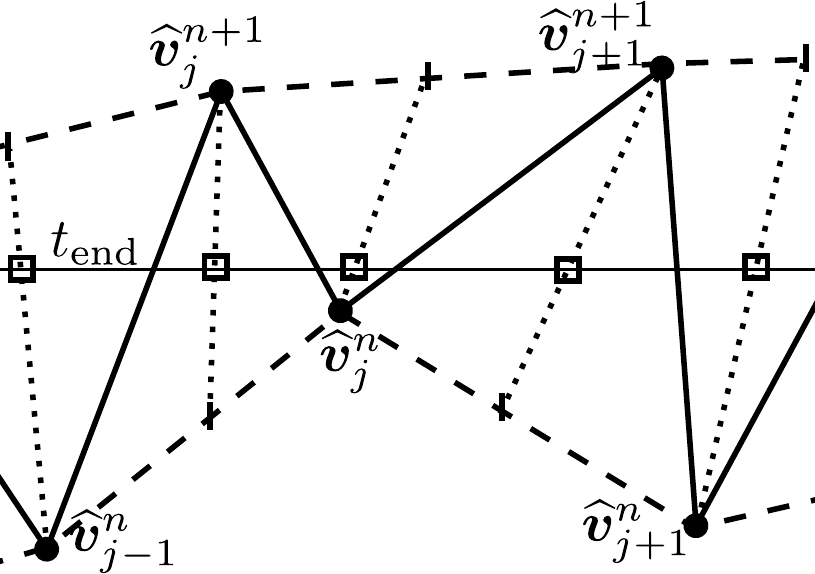}%
\caption{Triangular Interpolation.
An interpolation point is chosen for each triangle, placed in a centre location along the line $t=t\_{end}$.
Barycentric interpolation of the state $\bw$ is performed on each triangle.
}%
\label{fig:interpolation_triangular}%
\end{figure}

The free scattering of intersection nodes makes data points irregular and unsynchronised, depending on the system being simulated.
Scattered point interpolation, described in \autoref{fig:interpolation_triangular}, is used to determine end states.

The strength of the \MOSC{}  is that it entails no numerical diffusion or state interpolation other than that used for the end states.
It also automatically provides the longest allowable time step for each point individually. 
Weaknesses are the lack of conservation, that the method does not support fixed mesh arrangements,  requires end state interpolation and is poorly suited for handling shock discontinuities.

\subsection{A Method of Interpolated Point Characteristics (\MOPIC{})}
\label{sec:MOPIC}
\begin{figure}[h!ptb]%
\includegraphics[width=\columnwidth]{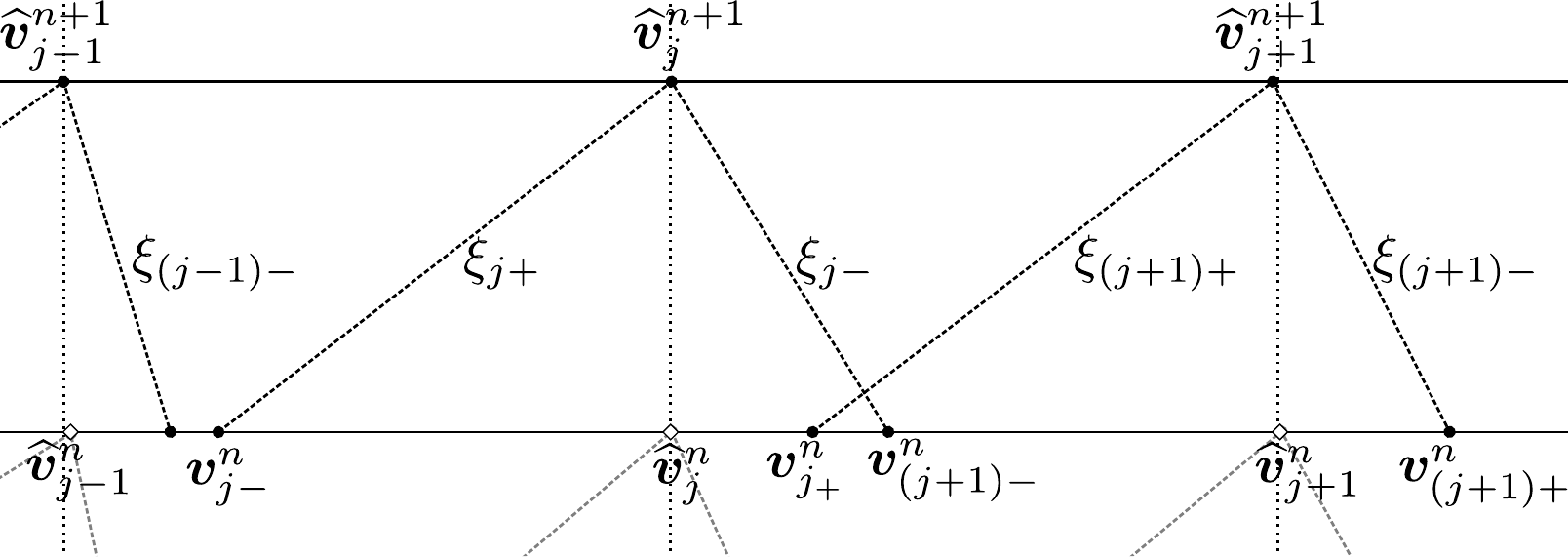}%
\caption{ Sketch of the \MOPIC{}.}%
\label{fig:MOPIC}%
\end{figure}

This method is similar to the characteristics method used by \egg\ Bournaski in \cite{Bournaski_MOC_dispersed_virtual_mass} for simulating dispersed pipe flow with a virtual mass effect.
The method of point interpolated characteristics, abbreviated `\MOPIC{}', allows for a fixed-mesh uniform time-step formulation of the \MOSC{}  through the use of spatial interpolation.
\autoref{fig:MOPIC} gives an illustration. 
Equal node spacing will here be applied.
Linear spatial interpolation of $\{\mmhat\bw_\j\n\}$ 
\begin{align*}
\mmhat\bw\jpm &= \mhat\bw_\j - \br{\mhat\bw_\j-\mhat\bw\jm}{\dthat\n \lambda\jpm\lpm}/{\dx},  &   \lambda\jpm\lpm &>0 \\
\mmhat\bw\jpm &= \mhat\bw_\j - \br{\mhat\bw\jp-\mhat\bw_\j}{\dthat\n \lambda\jpm\lpm}/{\dx},  &   \lambda\jpm\lpm &<0 
\end{align*}
is applied at each time level to compute point states $\mhat\bw\jpm\n$, which are integrated up to the new time level using \eqref{eq:MOC_end_eq}.
Higher order interpolation would not necessarily improve the scheme accuracy as this may violate the MOC's domain of dependence \cite{Suwan_MOL}.
Including interpolated states in an iteration process for the intermediate state does not seem to be a stable procedure. 
Instead, $\mmhat \phi\jpm\nh=\phi\of{\h\big(\mhat \bw_\j\n+\mhat \bw_\j\nn\big)}$ is used for the intermediate integrands.
Time steps are chosen
$
\dthat\n = \CFL\,\dx/\max\{|\lambda\jpm\lpm|\},
$
$\CFL$ being the \Courant{} number.

Also the \MOPIC{} lacks  the conservation property.

\subsection{A Method of Interpolated Cell Characteristics (\MOCIC{})}
\label{sec:MOCIC}
\begin{figure}[h!ptb]%
\includegraphics[width=\columnwidth]{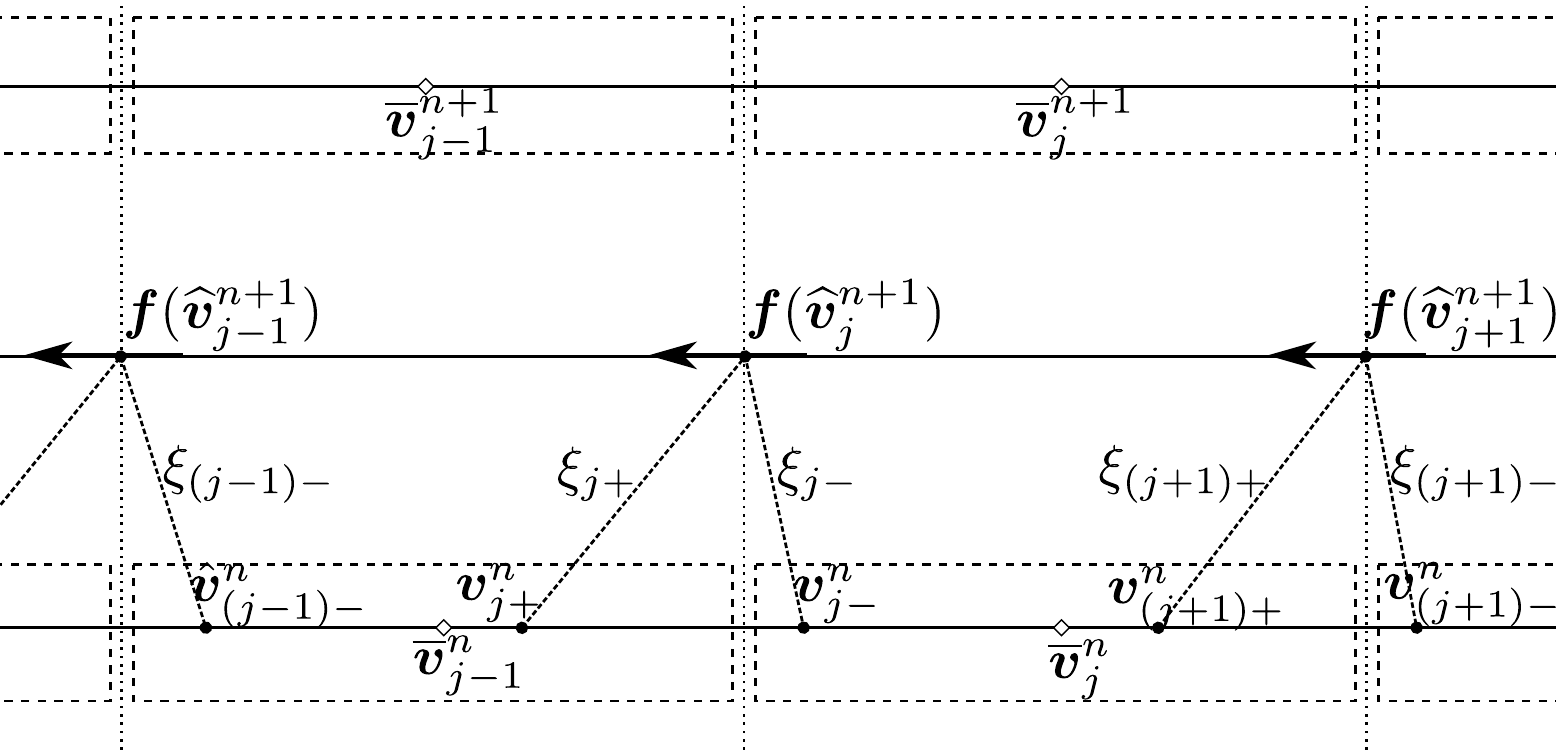}%
\caption{ Sketch of the \MOCIC{}.}%
\label{fig:MOCIC}%
\end{figure}

A new hybridisation with a finite volume method is now proposed with the intention of achieving conservation.
The \MOCIC{} 
is a combination of the \MOPIC{} with the finite volume method on a fixed grid.
\autoref{fig:MOCIC} shows an illustration.
Instead of proceeding from the states $\{\mhat \bw_\j\nn\}$ computed from the characteristic equation~ \eqref{eq:MOC_end_eq}, these intersection states are used to compute the fluxes of a fixed grid finite volume method. 
Adopting Eulerian time integration on the source term $\an {\ol\bs}\n\approx \bs\of{\ol\bw\n} =\ol\bs\n$ and intermediate fluxes $\an \bff \n\approx \bff\nh$ renders the finite volume equation \eqref{eq:avg_Holmaas} 
\begin{equation}
\ol\bw_\J\nn = \ol\bw_\J\n - \frac{\dt\n}{\dx_\J}\br{\bff\Jph\nh-\bff\Jmh\nh} + \dt\n \ol\bs_\J\n.
\label{eq:fluxes_fixed_schemes}
\end{equation}
Characteristic intersection points are used for the intermediate fluxes
\[
	\bff\Jmh\nh = \bff\of{\mhat \bw_\j\nn},
\]
where $\mhat \bw_\j\nn$ is computed from \eqref{eq:MOC_end_eq} in an interpolated fashion similar to that of the \MOPIC{}. Cell averages will in the point interpolation of $\mhat \bw\jpm\n$ be equivalent to centre point values and a uniform grid is used:
\begin{equation*}
\mmhat \bw\jpm\n = \tfrac12\br{\ol\bw_\J\n + \ol\bw\Jm\n} - \br{\ol\bw_\J\n - \ol\bw\Jm\n} {\dthat\n \lambda\jpm\lpm} / {\dx}.
\end{equation*}
A stable choice of intermediate integrands is $\mmhat\phi\jpm\n = \phi\of{\mhat\bw_\j\nn}$.
Time steps of the characteristic intersections are 
$
	\dthat\n = \h\dx/\max\{|\lambda\jpm\lpm|\}
$
and for the finite volume method 
$
	\dt\n =  2\, \CFL{} \dthat\n =  \CFL\,\dx/\max\{|\lambda\jpm\lpm|\}
$.

The \MOCIC{} is conservative at the expense of whatever numerical diffusion is inherent in the finite volume representation. Spatial interpolation errors are still present.

\subsection{A Method of Cell Centred Characteristics (\MOCCC{})}
\label{sec:MOCCC}
\begin{figure}[h!ptb]%
\centering
\includegraphics[width=1\columnwidth]{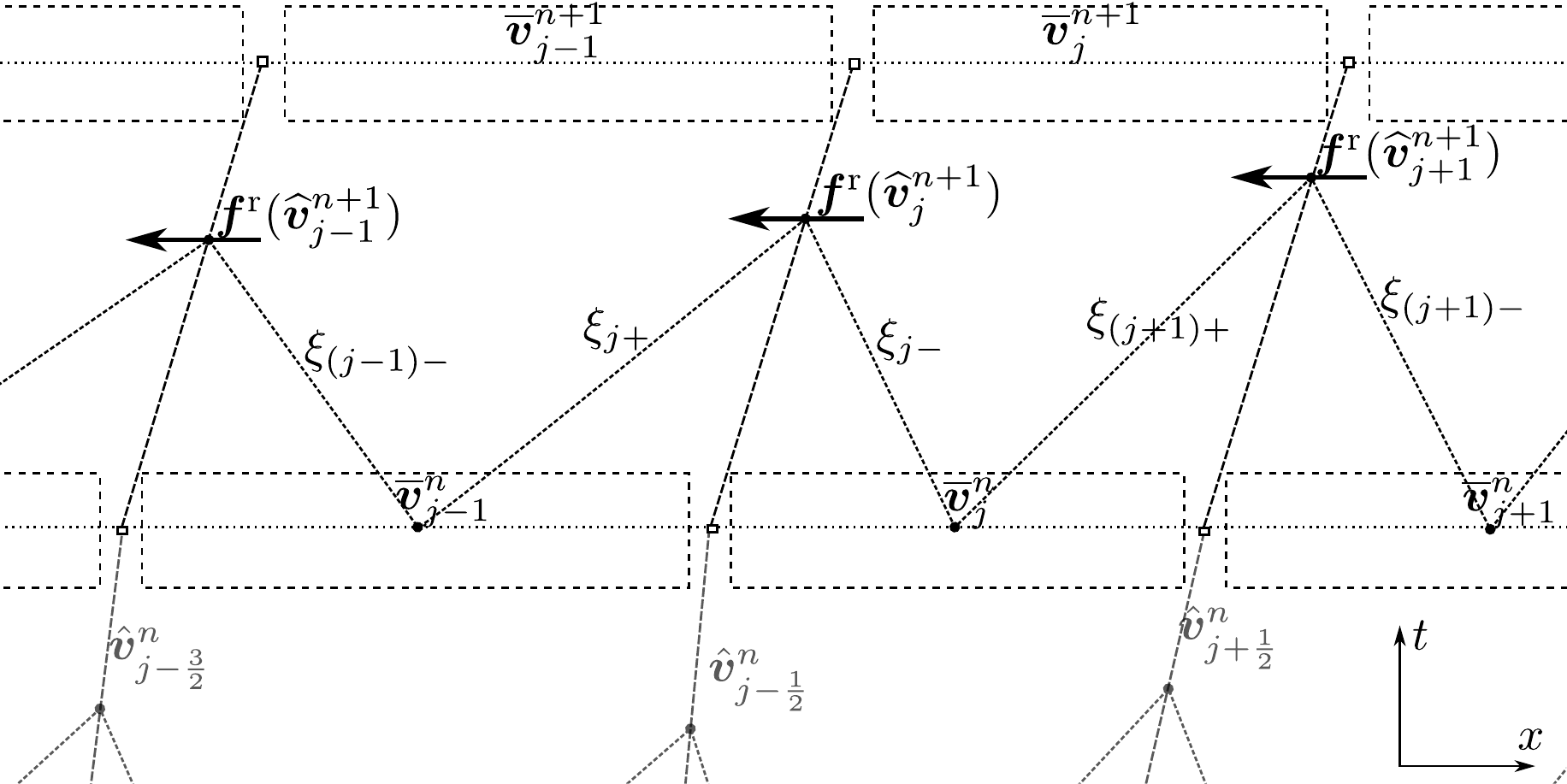}%
\caption{Illustration of the \MOCCC{}.}%
\label{fig:MOCCC}%
\end{figure}

The \MOCIC{} presented above can show signs of unexpected numerical instability, most 
{predominant} is the rarefaction wave of the Riemann problems in \autoref{sec:exp:breaking_dam}. 
Errors from the spatial interpolation are likely sources for this type of instability. 
A new proposition,
The Method of Cell Centred Characteristics (\MOCCC{}), removes the need for spatial interpolation by abandoning the spatially fixed grid. 
\autoref{fig:MOCCC} illustrates the principle.
The method is similar to the \MOCIC{}, but instead of following the linearised characteristic paths backwards from a pre-determined intersection point $(\mhat x,\mhat t)_\j\nn$ to some point of interpolation $(\mhat x,\mhat t)\jpm\n$, the characteristic paths are followed forwards in time from cell centre positions $(x,t)_\J\n$ to the new intersection points, as in the \MOSC{}. 
Intermediate integrands are computed $\mhat\phi\jpm\nh = \phi\of{\h \big(\mhat\bw\jpm\n + \mhat \bw_\j\nn\big)}$. 
The cell centre states are here taken as cell averages, \ie, the characteristic paths are integrated from $\mhat\bw_{\j\udd+}\n = \ol\bw\Jm\n$ and $\mhat\bw_{\j\udd-}\n = \ol\bw_\J\n$.

Cell face translation velocities are then 
\begin{equation}
\cc\Jmh\n = \frac{\mhat x_\j\nn-x\Jmh\n}{\mhat t_\j\nn-t\n}
= \frac{\dx_\J\n \lambda_{\j\udd+}\lp  +  \dx\Jm\n \lambda_{\j\udd-}\lm}{\dx_\J\n  +  \dx\Jm\n }
\label{eq:MOCCC:c_j}
\end{equation}
and the next cell face positions are
\begin{equation}
x\Jmh\nn = x\Jmh\n+\cc\Jmh\n \dt\n.
\label{eq:MOCCC:x_nn}
\end{equation}
The translation velocity ensures that the cell faces pass through the new characteristic intersection point.
Relative flux averages are computed from the characteristic intersection as an intermediate state $\an{\rel \bff}\Jmh\n = \rel \bff\of{\mhat\bw_\j\nn}$ and $\br{\dx \ol \bw}_\J\nn$ is then given from \eqref{eq:avg_Holmaas}. Eulerian time integration $\an {\ol\bs}\n\approx \bs\of{\ol\bw\n}$ is again applied for the source term,
and the averages states $\ol\bw_\J\nn = (\dx\ol\bw)_\J\nn/\dx_\J\nn$ are computed after translation.
Time steps are chosen 
$
\dt\n =  \CFL\cdot 2\, \min_\j \dthat_\j\n
= \CFL\cdot \min_\j (\dx_\j\n + \dx\jm\n)/(\lambda_{\j\udd+}\lp-\lambda_{\j\udd-}\lm).
$
\\

Negative cell lengths are very unlikely to occur in a simulation of natural flow. 
Expressions \eqref{eq:MOCCC:c_j} and \eqref{eq:MOCCC:x_nn}, alternatively studying \autoref{fig:MOCCC}, reveals that cell length irregularities are intrinsically counteracted by the cell propagation set-up ($\dx_\J\n \rightarrow 0$ guarantees ${\dx_\J\nn > 0}$.)
Negative cell length can only occur if integrated in time directly from a long cell (as compared to the neighbouring cells) 
beyond the time step of the characteristic intersection.
The characteristics paths would then have to be very irregular. 
A formal guarantee against negative cell lengths can be provided by imposing
$
	\dt < \min\limits_\J\wigbrac{ \frac{\dx_\J}{\cc\Jmh-\cc\Jph}\big|\; \J: \cc\Jmh>\cc\Jph}.
$
The numerical examples presented here never come close to activating this limit.

\section{A Roe Scheme}
\label{sec:Roe}
Roe's approximate Riemann solver \cite{Roe_original} is among the most popular finite volume schemes for non-linear hyperbolic conservation laws. 
Its main principle lies in solving linearised Riemann problems
\begin{equation}
\begin{gathered}
\bw_t + \hat \Jac\of{\bw\JL,\bw\JR } \bw_x = 0
\\
\bw\of{x,0} = \bw\JL\; (x<0),\quad \bw\of{x,0} = \bw\JR\; (x>0)
\end{gathered}
\label{eq:Roe_problem}
\end{equation}
at the cell faces and time step. $\hat \Jac$, $\bw\JL$ and $\bw\JR$ are constants respective to each cell face. 
Roe schemes are effective at discontinuities, but they require the formulation of so-called Roe-averaged matrices $\hat \Jac\of{\bw\JL,\bw\JR}$ at the cell faces with the  properties that
\begin{enumerate}[i)]
	\item \label{en:Roe:diag} $\hat \Jac\of{\bw\JL,\bw\JR}$ is diagonalizable with real eigenvalues,
	\item \label{en:Roe:consistant} $\hat \Jac\of{\bw\JL,\bw\JR } \rightarrow \Jac\of{\bw}$ smoothly as $\bw\JL,\bw\JR \rightarrow\bw$ and
	\item \label{en:Roe:main} $\hat \Jac\of{\bw\JL,\bw\JR} \diffRL{\bw} = \diffRL\bff$
\end{enumerate}
where \[\diffRL\cdot=\br\cdot\JR-\br\cdot\JL.\]
Generally, $\phi\JL = \phi\of{\bw\JL}$ and  $\phi\JR = \phi\of{\bw\JR}$.
The first and second properties are required for hyperbolicity and consistency, respectively. 
The third property ensures, by the Rankine-Hugoniot condition, that single shocks of the linear system \eqref{eq:Roe_problem} are shocks of the non-linear system~\eqref{eq:base_model}.

Consider the following splitting of the flux function:
\begin{equation}
\bff = \bfu + \bfh
\label{eq:bff_split}
\end{equation}
where
\begin{align*}
\bfu &= 
\begin{pmatrix}
	\al \ul \\ \h \rhol \ul^2-\h \rhog \ug^2
\end{pmatrix},
&
\bfh &=
\begin{pmatrix}
	0 \\  \my h
\end{pmatrix}. 
\end{align*}
We need a suitable integration path over which $\bff$ is easily evaluated; $\bff$ is written in terms of a parameter vector $\bwp$, rendering it  a low-order polynomial. Primitive variables are suitable in the case of $\bfu$, \ie
\begin{equation*}
\bwp = \br{\al,\ag,u\l,u\g}^T.
\end{equation*}
Note that $\pdiff\bw\bwp$ is constant and that $\pdiff {\bfu}{\bwp}$ is linear in $\bwp$.
A linear path
\[
\bwp=\bwpt\of \ws = \bwp\_L+\diffRL\bwp \ws
\]
is chosen for the integration of $\bfu$. We get
\begin{align}
\diffRL {\bfu} 
&= \int\_L\^R\! \mr d\bfu 
= \int_0^1\! \pdiff {\bfu}{\bwp}\of{\bwpt\of{\ws}} \pdiff{\bwpt}{s} \,\mr d \ws
= \pdiff {\bfu}{\bwp}\of{\ol\bwp} \diffRL\bwp \nonumber
\\
&= \pdiff {\bfu}{\bw}\of{\ol\bwp} \pdiff {\bw}{\bwp} \diffRL\bwp
= \Jacu\of{\ol\bwp} \diffRL\bw,
\label{eq:bff_ast}
\end{align}
{where $\ol\bwp = \tfrac12(\bwp\_R+\bwp\_L)$.}
The fourth expression is a result of $\pdiff {\bfu}{\bwp}$ being linear in $\bwp$, and the fifth and sixth from $\pdiff\bw\bwp$ being constant.
$\Jacu = \pdiff{\bfu}\bw$ is the Jacobian of $\bfu$, equalling \eqref{eq:Jac} without the 
$\br{\frac{\rho\l}{\al}+\frac{\rho\g}{\ag}} \my\dHofal$
term.

Consider now $\bfh$. We write
\begin{align}
\diffRL{\bfh} 
= \br{0, \my \diffRL h}^T
= \Jach\of{\a\_{\ell,L},\a\_{\ell,R}}\diffRL\bw,
\label{eq:bff_apo}
\end{align}
where
\begin{equation*}
\Jach = 
\begin{pmatrix}
0&0\\ \my {\diffRL h}/{\diffRL \al} &0
\end{pmatrix}.
\end{equation*}
Inserting \eqref{eq:bff_ast} and \eqref{eq:bff_apo} into \eqref{eq:bff_split},
\begin{equation*}
\diffRL \bff = \diffRL {\bfu} + \diffRL {\bfh} = \br{\Jacu + \Jach}\diffRL\bw,
\end{equation*}
the Roe average matrix
\begin{equation}
\hat \Jac = \Jacu\of{\ol\bwp} + \Jach\of{\a\_{\ell,L},\a\_{\ell,R}}
\label{eq:Roe_matrix}
\end{equation}
is seen to be the Jacobian \eqref{eq:Jac} constructed from arithmetically averaged primitive variables $\ol \ak$ and $\ol {u\k}$, with ${\diffRL h}/{\diffRL \al}$ replacing $\dHofal$. We use $\dHofal\of {\ol \al} $ close to $\diffRL \al = 0$ to avoid numerical 0/0-issues.

Once $\hat \Jac$ is formulated
\begin{equation}
\bff\of{0,t}= \tfrac12\br{\bff\JR+\bff\JL} -\tfrac12 \big|\hat \Jac\big| \diffRL{\bw}
\label{eq:Roe_scheme}
\end{equation}
provides the solution of the linearised problem \eqref{eq:Roe_problem}.
Here, 
\begin{align}
\big|\hat \Jac\big| &= \hat \LL\inv 
\big|\hat\Lamb\big|
\hat\LL 
=\begin{pmatrix}
	\big|\hat\lambda^+\big|+\big|\hat\lambda^-\big| 			& \br{\big|\hat\lambda^+\big|-\big|\hat\lambda^-\big|}\big/\hat\kappa \\
	\br{\big|\hat\lambda^+\big|-|\hat\lambda^-\big|}\hat\kappa  &\big|\hat\lambda^+\big|+\big|\hat\lambda^-\big|  
\end{pmatrix}
\label{eq:abs_Roe_matrix}
\end{align}
with the `hat' indicating the Roe intermediate state which in \eqref{eq:Roe_matrix} is the state of arithmetically averaged primitive variables and $ {\diffRL h}/{\diffRL \al}$ replacing $\dHofal$.

The solution \eqref{eq:Roe_scheme} is applied for each cell flux $\bff\Jmh$ in \eqref{eq:fluxes_fixed_schemes} without spatial reconstruction: $\bw\JR = \bw_\J$, $\bw\JL = \bw\Jm$.
Each time step is chosen 
$
\dt\n = \CFL\,\dx/\max_{\J,\pm}\big|\hat\lambda_{\J-\frac12}\pmn\big|.
$
The numerical tests presented herein are never in danger of promoting entropy violations in the Roe scheme, which may happen if an expansion fan straddles the time axis of problem~\eqref{eq:Roe_problem}. See \egg~\cite{LeVeque_2002_finite_volume_methods} for entropy corrections.


\section{Numerical experiments}
\label{sec:exp}

Those schemes not based on the finite volume averages, the \MOSC{}  and the \MOPIC{}, are generally not conservative and will slowly lose or gain mass and momentum depending on numerical errors. The loss of momentum is slow enough for the source term to counteract, but changes in the total liquid amount will become noticeable in long running simulations.  
These errors are suppressed in the linear stability tests of \autoref{sec:exp:VKH} and roll-wave tests of \autoref{sec:exp:roll_wave} by uniformly distributing the phase fraction error during runtime.
This is done in the manner $\alj\n := \alj\n + \ol{\alj^0}-\ol{\alj^n}$, bars here indicating the spatial average over the entire pipe. 
\\

Simulation results will also be compared with a more basic method to provide a better perspective. 
The staggered `donor-cell' or `upwind' scheme, here abbreviated `SUW', is still commonly used for the two fluid model \cite{Issa_two_phase_capturing, Bonizzi_MAST_2009,Kjeldby_capturing_vs_init_crit}.
It is formulated on a staggered grid stencil, collecting information from the direction of convection.
For completeness, this scheme is presented in Equation~\eqref{eq:SUW} assuming convection from left to right.
\begin{subequations}%
\begin{align}
\a_{\ell,J}\nn &= \a_{\ell,J}\n - \frac{\dt}{\dx}\br{ \a_{\ell,J}\n u_{\ell,\j+1}\n - \a_{\ell,J-1}\n u_{\ell,\j}\n  }, \\
\nonumber \q_\j\nn &= \q_\j\n 
- \frac12\frac{\dt}{\dx} \sqbrac{
 \br{\rhol\ul^2-\rhog\ug^2 }_j\n  - \br{\rhol\ul^2-\rhog\ug^2 }_{j-1}\n}  \\
	 &\quad - \frac{\dt}{\dx}\my\br{h_{J}\nn - h_{J-1}\nn  }+ \dt\: s_\q\big(\msmaller{ \a_{\ell,J}\nn,\q_\j\n}\big).
\end{align}%
\label{eq:SUW}%
\end{subequations}%
{Characteristic speeds are not computed in this type of scheme; following Liao et.\ al.}~\cite{Liao_von_Neumann} {the time step is based on the liquid velocity $\dt = \CFL \ul/\dx$.}
New area fraction information is used in the momentum equation after first solving the mass equation. 
This makes the scheme more implicit and numerically stable, but also more diffusive. 
Alternatively, measures like selecting smaller time steps, using a non-staggered grid, etc.\ also stabilise the scheme, though
all such options increase stability at a higher cost of increased numerical diffusivity. 

This staggered upwind scheme provides less numerical diffusion than its non-staggered equivalent, which in turn is less diffusive than the Lax-Friedrich scheme, results from which are not shown for the sake of briefness.

\subsection{A Riemann problems}
\label{sec:exp:breaking_dam}
A 
{Riemann problem is} here presented as a first test, the results of which are also relatable to surge wave cases and similar. 
Initial conditions $\bw\of{x<0, 0}=\bw\JL$, $\bw\of{x>0, 0}=\bw\JR$ are 
{$u\_L = \unitfrac[1.25]ms$, $u\_R = \unitfrac[0.75]ms$, $h\_L = .75\D$, $h\_R = .25\D$} and $\theta = \rho\g = \tau\g=\tau\l = 0$, \ie, an inviscid free-surface flow.
A channel geometry is imposed by temporarily defining $\Hofal = \al/\D$. 
{The two-fluid model is equivalent to the shallow water equations under these conditions.
$100$ nodes/cells used in the domain $-\D/2\leq x\leq \D/2$.}
The left and right velocities have been chosen high enough for the flow to be supercritical ($\lambda\ud+, \lambda\ud- >0$); 
the primitive upwind scheme becomes unstable at the shock if the flow is sub-critical (regardless of time step.) 
Same applies for primitive non-staggered upwind schemes and the Lax-Friedrich scheme.
{\Courant{} numbers are chosen small enough to avoid numerical oscillations at the shock and are listed in \autoref{tab:CLF}.}
\\

\begin{figure}[h!ptb]%
\begin{subfigure}{\columnwidth}
\includegraphics[width=\columnwidth]{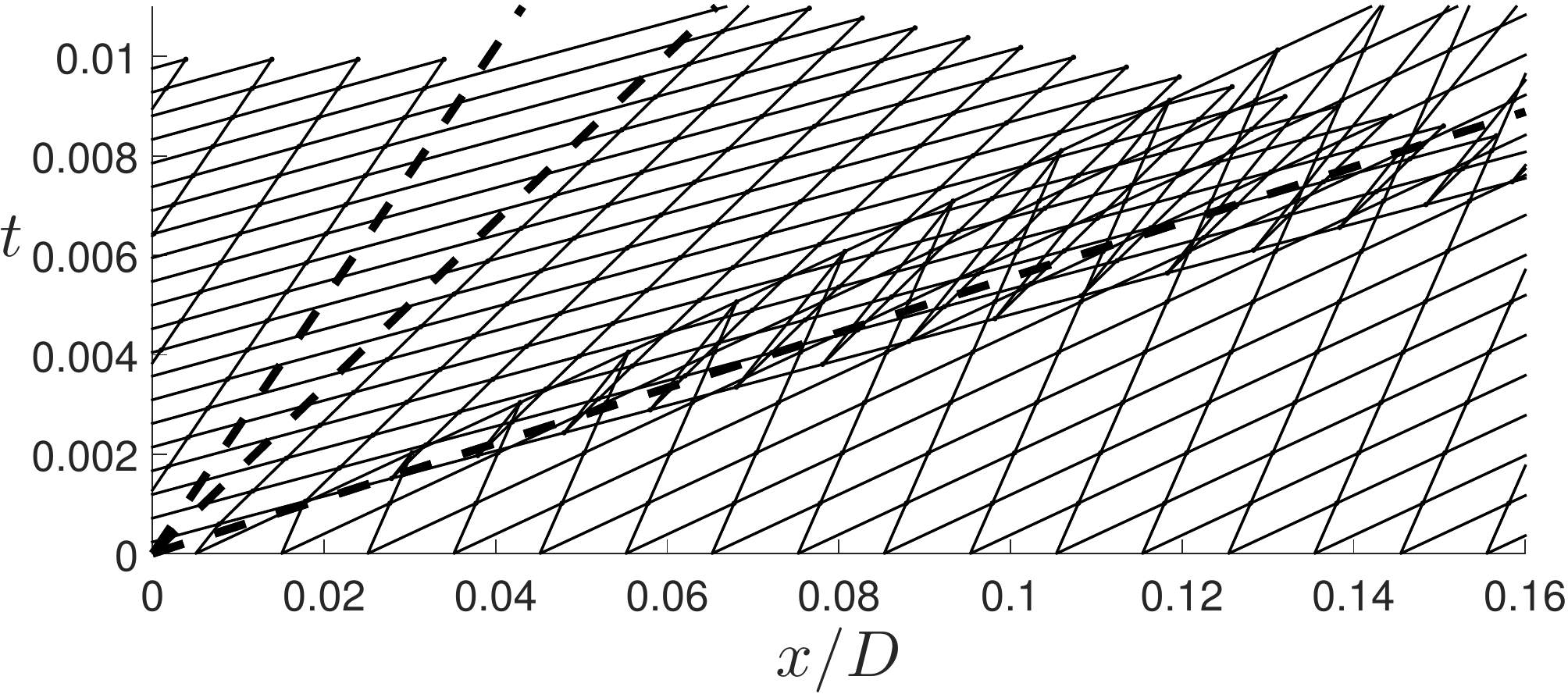}
\caption{Without shock conditioning}%
\label{fig:exp:breaking_dam_MOCspacetime:noshockfix}%
\end{subfigure}
\begin{subfigure}{\columnwidth}
\includegraphics[width=\columnwidth]{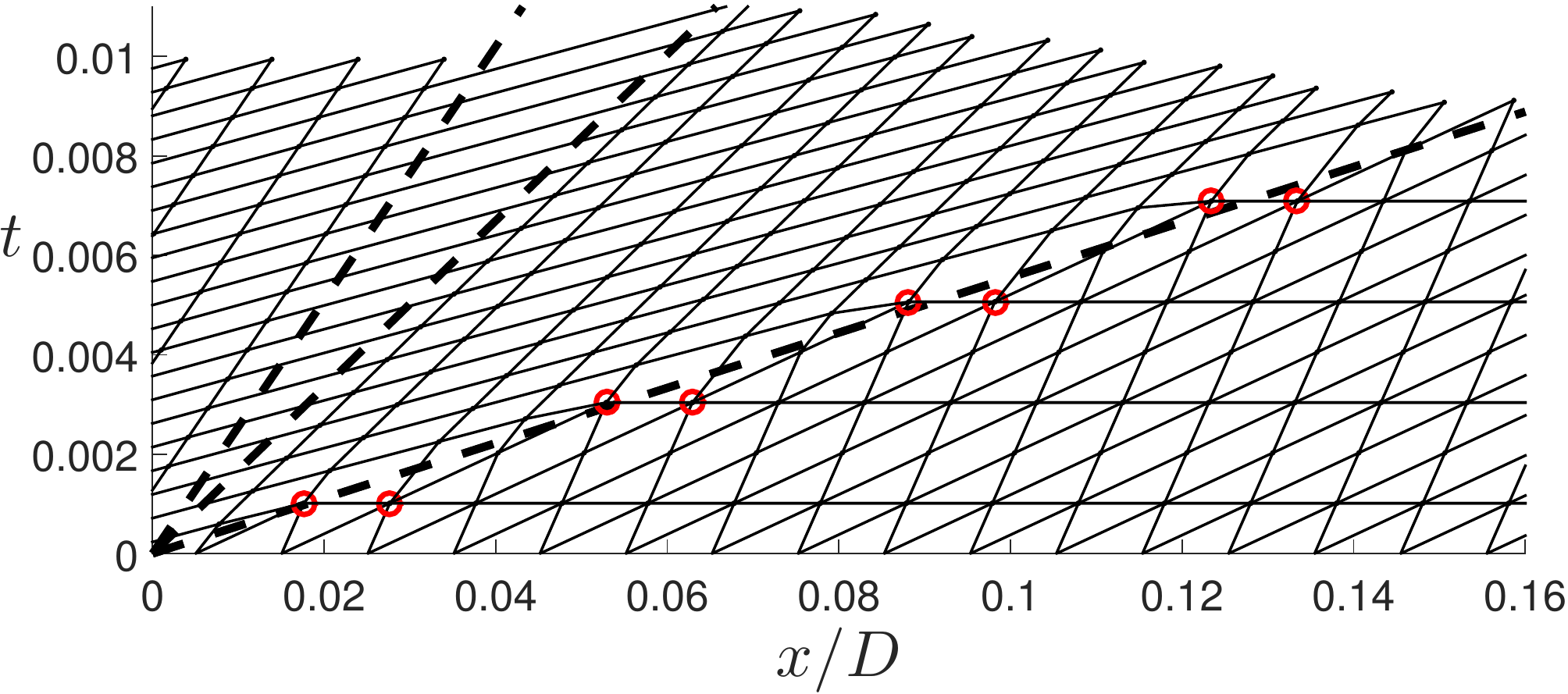}
\caption{With shock conditioning}%
\label{fig:exp:breaking_dam_MOCspacetime:shockfix}%
\end{subfigure}
\caption{\MOSC{} node path cf.\ \autoref{fig:exp:breaking_dam}.}%
\label{fig:exp:breaking_dam_MOCspacetime}%
\end{figure}

Consider first the \MOSC{}.
Shocks will, in the \MOSC{}, manifest as the clustering of characteristic points in space and time.
Nodes downstream a shock will progress quicker in time than those upstream, such that the present characteristic of one node eventually crosses an older node path computed many time steps previous.
This results is an inversed fan of ambiguous, overlapping states.
\autoref{fig:exp:breaking_dam_MOCspacetime:noshockfix} shows such an inversed fan.
A shock conditioning routine is imposed on the \MOSC{} 
which enables 
the simulation to proceed without node paths crossing each other
-- \autoref{fig:exp:breaking_dam_MOCspacetime:shockfix}.
This routine consists of occasionally excluding progressed nodes from the time integration step and removing node pairs whose paths 
{would otherwise cross in a manner not handled by the scheme.}
{Shock fitting approaches, such as described in \cite{Salas_shock_fitting}, may be possible, 
but are not pursued here.}
\\

\begin{figure}[h!ptb]%
\includegraphics[width=\columnwidth]{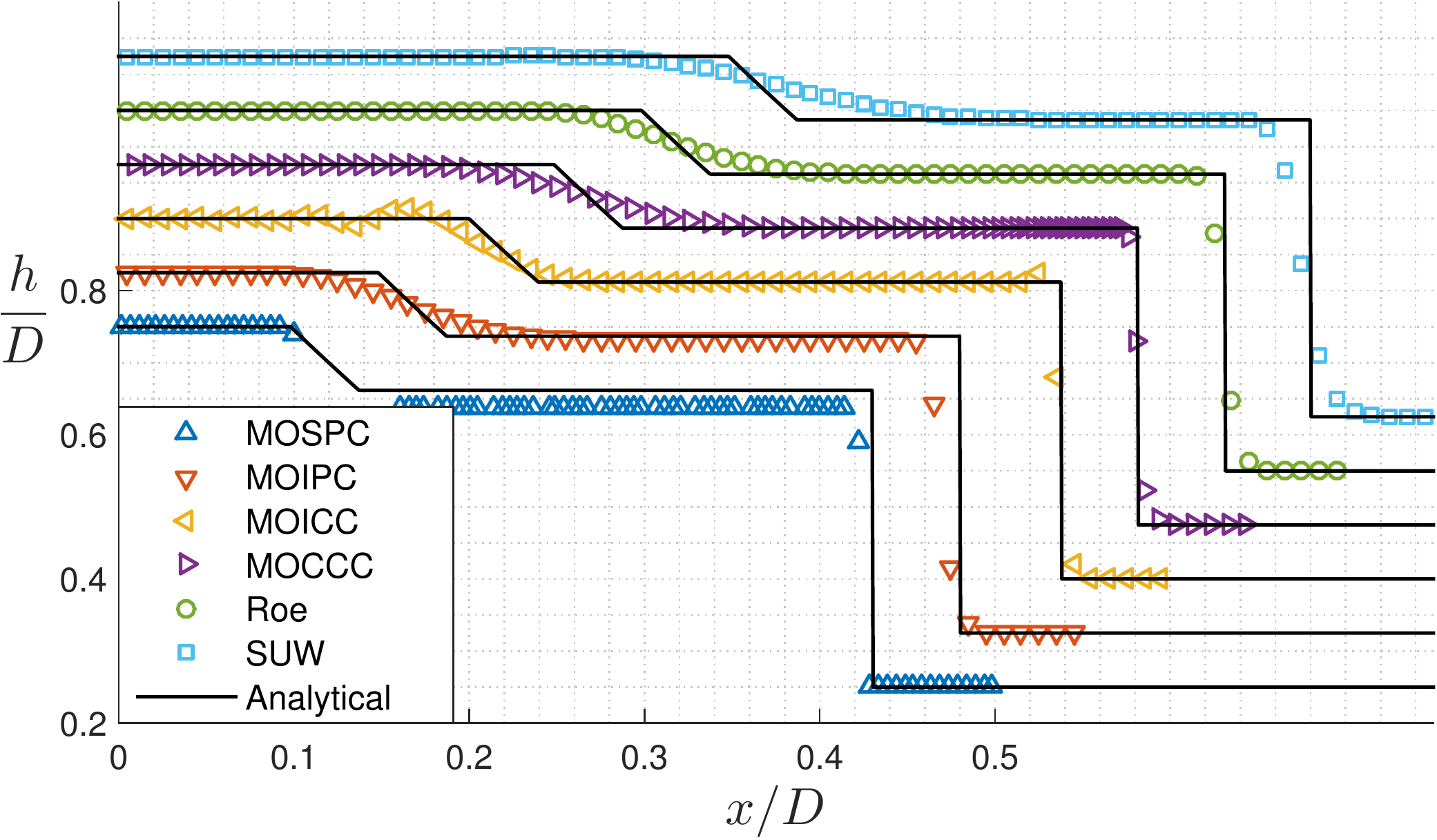}%
\caption{Level height/phase fraction at $t=\unit[0.025]s$ of a Riemann problem with initial discontinuity $u\_L = \unitfrac[1.25]ms$, $u\_R = \unitfrac[0.75]ms$, $h\_L = .75\D$, $h\_R = .25\D$ at $x=0$. $100$ nodes/cells used.}%
\label{fig:exp:breaking_dam}%
\end{figure}

\autoref{fig:exp:breaking_dam} shows the level height at $t=\unit[0.025]s$. 
{No diffusion is present in the \MOSC{} scheme.} 
As seen from \autoref{fig:exp:breaking_dam} and the \MOSC{}  space-time path plot of \autoref{fig:exp:breaking_dam_MOCspacetime:shockfix}, nodes bifurcate along the rarefaction wave wedge, splitting the left and right states in space. 
Linear interpolation within the rarefaction wave divide agrees with the  analytical solution.
{The intermediate state prediction is however not accurate due to the shock discontinuity.
(The `loose' point seen in the shock of the \MOSC{} simulation is from the end state triangular interpolation, the method itself experiencing no shock diffusion.) Also the \MOPIC{} scheme underpredicts the intermediate level height slightly because of lacking shock conservation.}
\\

The finite volume based schemes converge towards the analytical solution {if stable. 
They show minor numerical diffusion on the shock
and stronger diffusion at the rarefaction wave.
A numerical oscillation error is observed on both shock and rarefaction wave in the \MOCIC{} scheme.
Contrary to the trend in all the other schemes, these errors increase with decreasing \Courant{} numbers.
The non uniformity of the grid in the \MOCCC{} scheme is apparent close to the shock, where node are closely spaces.
This increases the sharpness of the shock front predicted by the \MOCCC{}, but it also reduces the allowable time step.
In fact, \autoref{tab:exp:dtdx} shows that the the time step benefit of moving grids is lost due to grid clustering in this problem due the the strong non-linearity.
The Roe scheme performs well, as expected on a Riemann problem.
}

\subsection{Linear instability}
\label{sec:exp:VKH}
{Viscous friction is included in the next two test cases.}
The viscous Kelvin-Helmholtz (VKH) criterion originates from applying linear stability theory to a uniformly stratified steady state solution -- see \egg, \cite{Barnea_stability_separated_flow}. 
The theory predicts flow \textit{instability}, with growing waves, \textit{if}
\begin{subequations}
\begin{equation}
	\my\dHofal - \rho\l  \br{U\l-\Ccrit}^2 - \rho\g  \br{U\g-\Ccrit}^2  < 0,
\label{eq:VKH:crit}
\end{equation}
where the critical perturbation celerity is 
\begin{equation}
	\Ccrit = \left.{\pdiff {\S_q} \al}\right/\!\br{ \pdiff {\S_q}{(\a u)\g} - \pdiff {\S_q}{(\a u)\l}}.
\label{eq:VKH:C}
\end{equation}
Upper-case symbols here indicate a steady state 
{and $\S_q$ is the `momentum' source term \eqref{eq:s} of gravity and friction, here formulated as a function of the volumetric flow rates.}
For a flow state to remain steady it must satisfy the so-called \textit{holdup equation}
\begin{equation}
	\S_\q\of{\Al,(A U)\l,(A U)\g} = 0.
\label{eq:VKH:holsup_eq}
\end{equation}%
\label{eq:VKH}%
\end{subequations}%
%
{
At the limit of neutral linear stability one can show from} \eqref{eq:VKH:crit} {and} \eqref{eq:lambda} {that the faster characteristic velocity equal the critical wave celerity $\Ccrit$, \ie, the flow is \textit{critical} relative to the perturbation also in a hydraulic sense.
}
Supercritical flow will in turn have to develop into a shock relative to the perturbation wave, telling us that the roll-wave regime will ensue if the pipe cross section is not breached first. 

%
\begin{table}[h!ptb]
\centering
\begin{tabular}{|rl|rl|}
\hline
liquid density&				$\rho\l$	&998&			$\unitfrac[]{kg}{m^3}$\\
gas density&				$\rho\g$	& 50&			$\unitfrac[]{kg}{m^3}$\\
liquid dynamic viscosity&	$\mu\l$ 	& 1.61\Em5&		$\unit{Pa\:s}$\\ 
gas dynamic viscosity&		$\mu\g$ 	& 1.00\Em3&	$\unit{Pa\:s}$ \\
internal pipe diameter&		$\D$ 		& 0.1&			$\unit[]m$	\\
wall roughness&							&2\Em5& 		$\unit[]m$ \\
Pipe inclination& 			$\theta$	& 1\degree&		$-$\\
Mean level height&			$\ol h$	& 0.02&				$\unit[]m$\\\hline
\end{tabular}
\caption{Fixed parameters for the tests in \autoref{sec:exp:VKH} and \ref{sec:exp:roll_wave}.}
\label{tab:parameters}
\end{table}

Fixed fluid and pipe properties are in the following tests the same as in \cite{Holmaas_roll_wave_model} for upwards-directed flow, presented in \autoref{tab:parameters}.
With these parameters, \eqref{eq:VKH} predicts that the stratified flow will turn unstable at $\Qmc/\area = \unitfrac[3.153]ms$. 
{
Scheme \Courant{} numbers have again been chosen based to the overall stability behaviour in uniform stratified, linear wavy and non-linear roll-wave flows have been considered
 -- see \autoref{tab:CLF}.
}
\\

\begin{figure}[h!ptb]%
\centering
\includegraphics[width=.85\columnwidth]{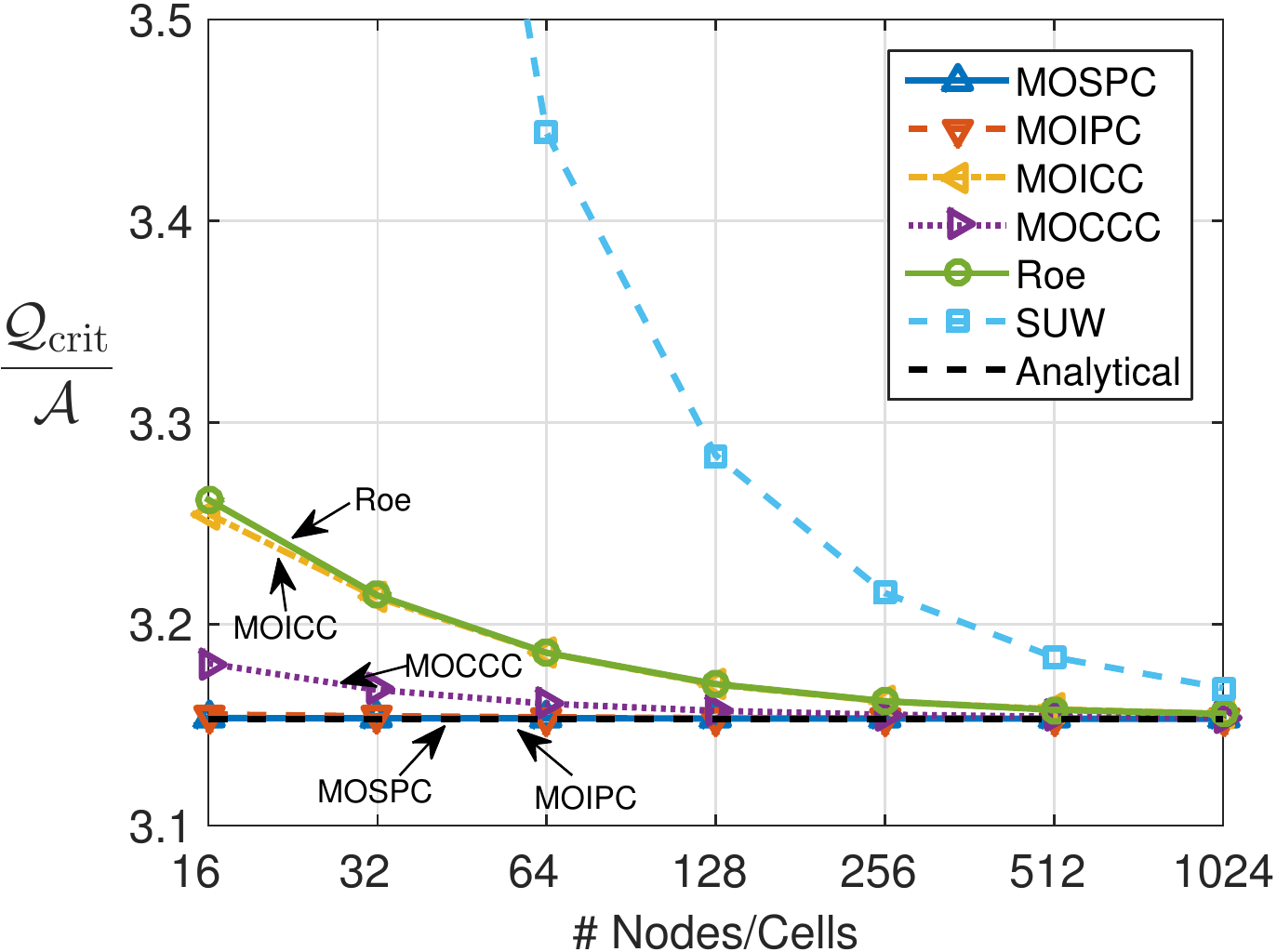}%
\caption{The critical mixture flow rate $\Qmc/\area$ above which the scheme predicts wave growth
Analytical solution from linear stability theory, Equations~\eqref{eq:VKH}.}%
\label{fig:exp:VKH}%
\end{figure}

\autoref{fig:exp:VKH} shows 
the observed values of the critical mixture velocity $\Qmc/\area$ at which wave growth is observed in a simulation. These values may be compared to the value from linear stability theory, labeled `Analytical' in the figure.
A disturbance of the order $1\Em7$ relative to the state variables were imposed on all simulations. 
The longest wavelength, spanning the entire simulation domain, is most resistant to numerical diffusion; simulations at the critical limit always grew into this wave, regardless of the disturbance.
\\

\begin{table}%
\centering
\begin{tabular}{|l|c|c|c|}\hline
&&\multicolumn{2}{c|}{$\CFL$ in Section\dots} \\
Scheme & $\dt = \dots$ & \ref{sec:exp:breaking_dam}& \ref{sec:exp:VKH}, \ref{sec:exp:roll_wave}\\[.25ex]\hline
\MOSC{} & natural intersection & $1.0$& $1.0$\\[.25ex]
\MOPIC{} & $\CFL{} \dx/\max{\{|\lambda\jpm|\}}$ &$1.0$ & $0.999$\\[.5ex]
\MOCIC{} & $\CFL{} \dx/\max{ \{|\lambda\jpm|\}}$ &$1.0$ & $0.95$\\[.5ex]
\MOCCC{} & $\CFL{} \min \frac{\dx_\j + \dx\jm}{\lambda_{\j\udd+}\lp-\lambda_{\j\udd-}\lm}$ &$0.6$& $0.95$\\[.5ex]
Roe & $\CFL{} \dx/\max{\big\{\big|\hat\lambda\jmh\big|\big\}}$ &$1.0$& $0.95$\\[.5ex]
SOU & $\CFL{} \dx/\max{\{|\ul|\}}$ &$0.3$& $0.5$\\\hline
\end{tabular}
\caption{Time step computation and \Courant{} numbers used in \autoref{sec:exp:breaking_dam} and \autoref{sec:exp:VKH} and \ref{sec:exp:roll_wave}.}
\label{tab:CLF}
\end{table}

The pure characteristic methods, the \MOSC{}  and the \MOPIC{}, show very precise stability predictions for all tested resolutions. 
The \MOCIC{}, \MOCCC{} and Roe schemes also show accurate predictions, whereas the staggered upwind scheme is dominated by numerical diffusion for the low-resolution simulations and fail to become unstable altogether in the 16 grid cell simulation.
A central feature of this test is that the error in $\Qmc$ is strongly dependent upon the choice of \Courant{} number. 
{The purely characteristic schemes, \MOSC{} and \MOPIC{}, perform perfectly.}
\\

The results of the other schemes seems very dependent upon the \Courant{} number, approaching the analytical solution as $\CFL$ nears unity. 
It seems therefore that the property of a method to remain stable at high \Courant{} numbers is strongly desirable in capturing wave instabilities.  
Finite volume schemes, excluding the staggered upwind scheme, where here all given the same \Courant{} number of $0.95$ based on an overall consideration of stability. 
A disadvantage of the staggered upwind scheme in this respect is that the time step limit is only estimated based on convective phase speeds rather than the characteristic speeds.

\subsection{Roll-wave}
\label{sec:exp:roll_wave}
\begin{figure}[h!ptb]%
\includegraphics[width=\columnwidth]{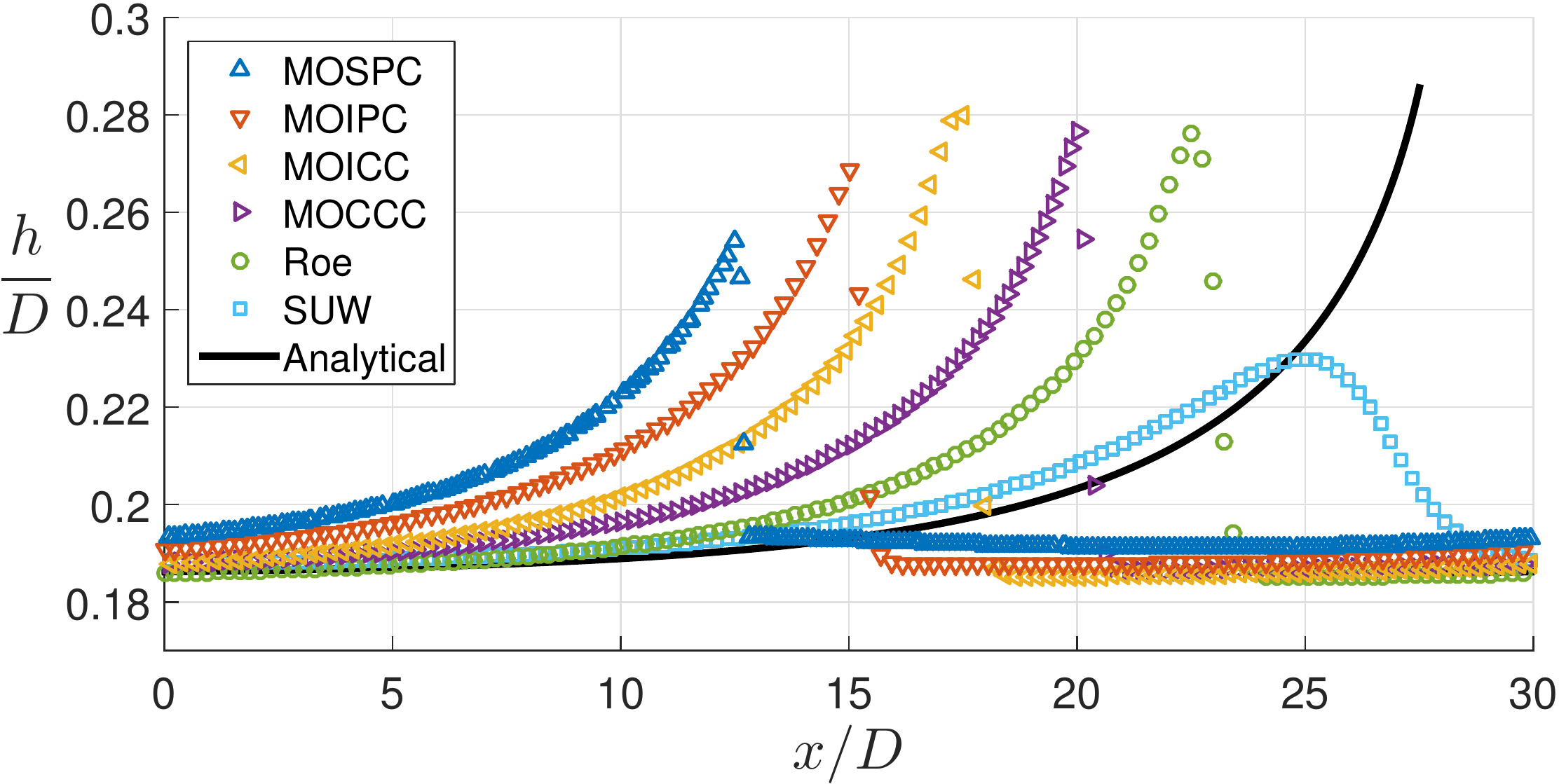}%
\caption{Single wave simulations, steady state, 128 nodes/grid cells. $\Qm/\area = \unitfrac[3.4]ms$  and the parameters in \autoref{tab:parameters}.}
\label{fig:exp:MOCs_Roe_UWS_Nj128}%
\end{figure}

\begin{figure}[h!ptb]%
\centering
	\begin{subfigure}{.8\columnwidth}
	\includegraphics[width=\columnwidth]{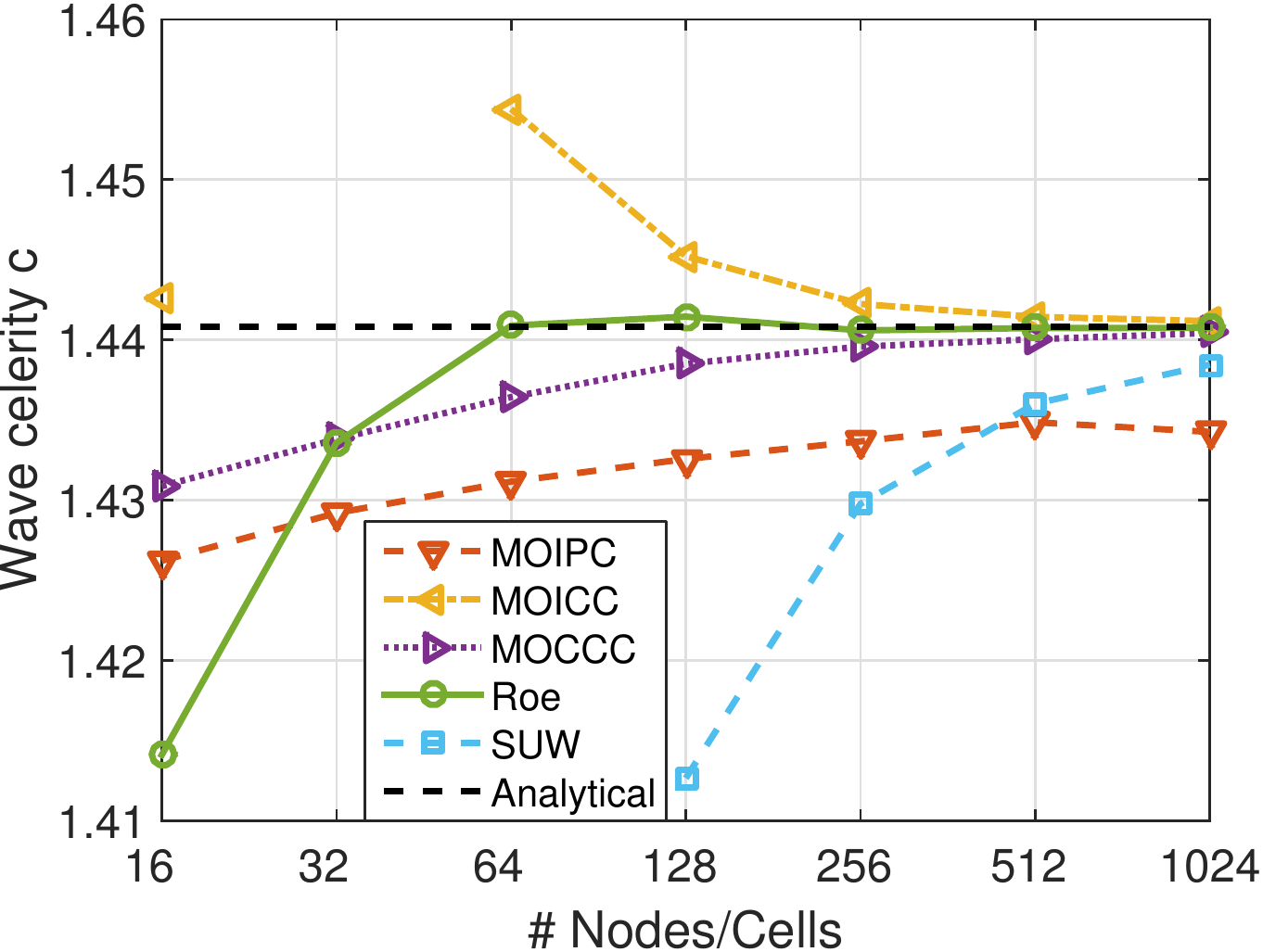}%
	\caption{Wave celerity. Analytical celerity: $1.4408 \unitfrac ms$}%
	\label{fig:exp:rollwave:wavespeed}%
	\end{subfigure}
	\\
	\begin{subfigure}{.8\columnwidth}
	\includegraphics[width=\columnwidth]{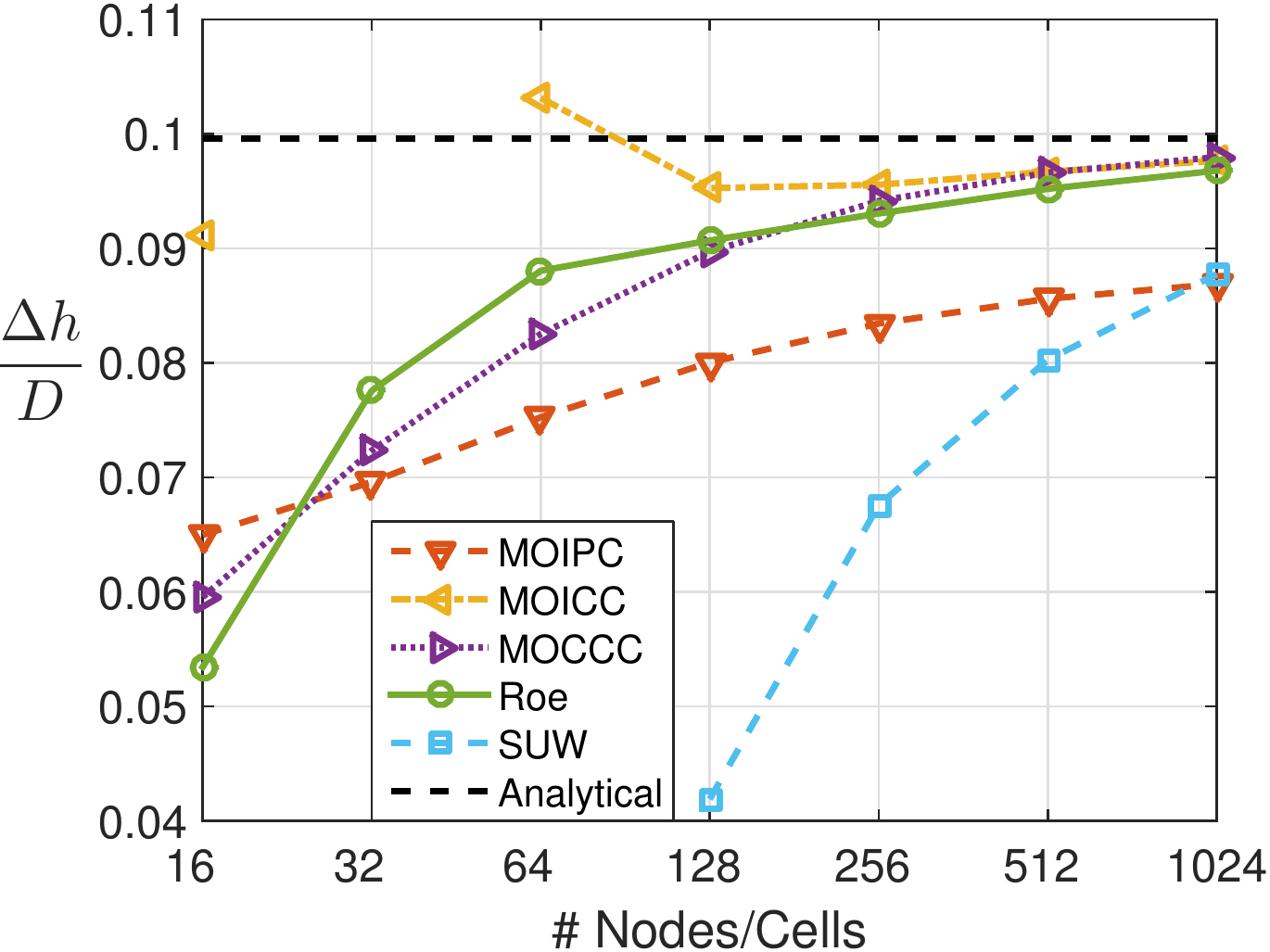}%
	\caption{Wave height. Analytical wave height: $0.0996 \, d$}%
	\label{fig:exp:rollwave:waveheight}%
	\end{subfigure}
	%
	%
\caption{Single wave simulation. $\Qm/\area = \unitfrac[3.4]ms$ and the parameters in \autoref{tab:parameters}.}%
\label{fig:exp:rollwave}%
\end{figure}

Steady-state roll-wave predictions  are presented in this section. 
The same flow parameters and \Courant{} numbers are used here as in the linear stability test case of the previous section (\autoref{tab:parameters}.)
A higher mixture flow rate $\Qm/\area = \unitfrac[3.4]ms$ drives the wave regime. A small, single-period sinusoidal wave is used as initial condition and each simulation is run until a steady state is reached. (The term `steady state' refers here to flows which are time independent in a particular moving frame.)
These predicted profiles may be compared to the analytical, numerically integrated  profile solution of Watson~\cite{Watson_wavy}, labelled 'Analytical' in the plots.  
The wave celerity $c$ of this solution equals the wave front shock velocity and is obtained from the Rankine-Hugoniot condition that a shock invariant
\begin{equation}
\diffRL{ \tfrac12  \rhol (\ul-c)^2 - \tfrac12  \rhog (\ug-c)^2 +\my h } = 0
\label{eq:shock_invariant}
\end{equation}
should be maintained across the discontinuity.\footnote{
This can be seen from integrating \eqref{eq:base_model} in a relative frame thinly over the shock. All but the relative fluxes disappear.
} In this case, the wave celerity is $c = \unitfrac[1.4408]ms$.
\autoref{fig:exp:MOCs_Roe_UWS_Nj128} shows the predicted wave profiles in terms of level height for $128$ nodes/grid cells.
Wave celerity plots and wave height plots are presented in \autoref{fig:exp:rollwave} as functions of the node/cell number with a 2-based logarithmic abscissa. 
Displayed wave height values $\Delta h$ are the temporal averages $\langle\max_\J\{h_\J\n\}-\min_\J\{h_\J\n\}\rangle$ over a number of time steps after the respective waves reached a steady state. 
Likewise, the wave celerity values are similar time set averages $\big\langle \big(x_{\J\_{max}\nn}\nn-x_{\J\_{max}\n}\n\big)/\dt\n \big\rangle$, $\J\_{max}\n$ being the cell index whose liquid fraction is greatest at time level $n$.

The Roe scheme is seed to predict the wave celerity very well for $64$ grid cells and more. 
This is to be expected as the Roe scheme 
is a linearised Riemann solver.

The conservative characteristic methods, the \MOCIC{} and \MOCCC{}, are also seen to converge to the same celerity solution, though somewhat more slowly. 
Simulations turn numerically unstable with the \MOCIC{} at $32$ grid cells; a \Courant{} number of 0.95 does not appear to stabilise the \MOCIC{} scheme sufficiently in this case. 

The \MOPIC{} does not appear to converge towards the correct wave celerity. 
This is attributable to the methods' lack of conservation, particularly across the shock 
(continuity is indirectly assumed in the integration step \eqref{eq:diag_syst_int}-\eqref{eq:diag_syst_disc}.)

The \MOSC{}  also lacks the conservation property. 
In addition, the steady state solutions of the \MOSC{}  seems dependent upon whatever numerical trickery is applied for the shock conditioning. Lacking a well-defined  routine yielding consistent roll-wave results, the \MOSC{}  has been excluded from \autoref{fig:exp:rollwave}.
\\

\subsection{Time steps}
Finally, the method efficiencies in terms of time steps are considered. 
\autoref{tab:exp:dtdx} shows {normalised} average time steps 
$\an \dt$  computed as arithmetic averages of $\{\dt\n\}$.
$\an \dx$ is the domain length divided by the number of cells or nodes.
Normalised time steps were largely independent of the number of cells/nodes.
Where a slight dependency is present (foremost in the \MOCCC{}) the table shows the $128$ cell simulation value.
It is seen that the \MOCCC{} has about 4.6 times longer time steps that its fixed-grid counterparts in the plane stratified case and 3.2 times in the roll-wave case.
{In the Riemann problem the dynamic grid proved disadvantageous in regard do the time step lengths with a ratio around $0.6$ --
the time step advantage of moving grids is seen to diminish with increasing non-linearity due to cell clustering.}

\begin{table}%
	\centering
	\begin{tabular}{|l||c|c|c|}
	\hline
					&Riemann& VKH	& roll-wave \\\hline
	\MOSC{} &0.56 &5.8	& -- \\\hline 
	\MOPIC{}&0.45	&2.5	&2.2\\\hline 
	\MOCIC{}&0.46 &2.4	&2.1\\\hline
	\MOCCC{}&0.27 &10.9	&6.8\\\hline
	Roe			&0.46	&2.5	&2.1\\\hline
	SUW			&0.24	&1.7	&1.5\\\hline		
	\end{tabular}
\caption{Normalised mean time step $\an\dt {\mc U}/{\an \dx}$
where $\mc U=\frac 12 (u\_L+u\_R)$, $\Qmc/\area$ and $\Qm/\area$ for the Riemann, VKH	and roll-wave problem
of Section \ref{sec:exp:breaking_dam}, \ref{sec:exp:VKH} and \ref{sec:exp:roll_wave}, respectively. }
\label{tab:exp:dtdx}
\end{table}



\section{Discussion}
\label{sec:discussion}

One of the main arguments for resorting to characteristic methods is to detect the onset of hydrodynamic instability without  artificially stabilising the predictions with numerical diffusion. 
Artificial numerical stabilisation in staggered upwind schemes is already well documented \cite{Kjeldby_capturing_vs_init_crit, Liao_von_Neumann, Issa_two_phase_capturing}.
It would however appear that the numerical diffusion in the Roe scheme, with a \Courant{} number close to one, is small.
{In fact, testing suggests that linear VKH stability predictions provided by the simple first-order accurate Roe scheme proposed here converge towards the analytical solution as the \Courant{} number approaches unity --
a feature which will be investigated in the future.
Normally, however, the \Courant{} number must be subject to the requirements of global stability and cannot be optimised locally.
Still, a strategy where simple schemes adopt time steps suited to the wave propagation or characteristic information flow is an interesting alternative to methods focusing on state reconstruction and higher-order accuracy. 
}



\section{Conclusions}
\label{sec:conclusions}

Methods from the literature (\MOSC{}, \MOPIC{} and Roe's method) have been adopted to the incompressible two-fluid model for pipe flows.
In addition, hybrids of the characteristic and finite volume methods (\MOCIC{} and \MOCCC{}) have been proposed to achieve conservation.
The method of scattered point characteristics (\MOSC{}) is excellent for predicting the onset of viscous linear instability, even when the numerical resolution is poor. 
However, the method in this unrestricted form is poorly suited for the non-linear wave regime which follows. 
Better at handling the roll-wave regime are methods where nodes do not scatter and collide (\MOPIC{}) 
and schemes which are conservative (\MOCIC{}, \MOCCC{} and the finite volume methods.) 
Only the numerically conservative methods were observed to converge towards a correct roll-wave solution.
The Roe solver is designed for shock discontinuities and shows both better accuracy and stability in the presence of wave fronts. 
Though outperformed by characteristic methods, 
these finite volume methods also predict the onset of viscous linear instability quite well, even when poorly resolved, 
if allowed a \Courant{} number close to unity.
{Such \Courant{} numbers can only be chosen if waves with higher speeds are not present elsewhere in the flow. }

Method efficiency can be greatly enhanced if one allows for a dynamic grid arrangement, such that grid cells follow the main  drift in characteristic information. 
This seems a very natural part of characteristic methods, also enhancing the numerical stability and eliminating the need for interpolation.
The moving grid feature can also easily be expended to other finite volume methods.

On the whole, the Roe and the conservative versions of the MOC seem to yield results of similar quality. The characteristic methods show and advantage in the linear wave regime and the Roe scheme shows an advantage in the non-linear one. 
The Roe scheme also has the advantage that it 
{does not require a moving grid stencil, but can be adopted into one.}

\section*{Acknowledgements}
This work is financed by The Norwegian University of
Science and Technology (NTNU) as a contribution the Multiphase
Flow Assurance programme (FACE.)
The author would like to thank Tore Fl{\a}tten for his very useful feedback 
and Kontorbamse for the comforting support.

\input{nomenclature.sty}

{\small
\printnomenclature
}

\bibliographystyle{plainnat}
\bibliography{refs_PhD}

\end{document}